\documentclass[]{aa}
\usepackage{graphicx}
\usepackage{txfonts}
\usepackage{natbib}

\newcommand{\A}{\hbox{$\alpha$}}
\newcommand{\Ap}{\hbox{$\alpha^{'}$}}
\newcommand{\B}{\hbox{$\beta$}}
\newcommand{\Bp}{\hbox{$\beta^{'}$}}
\newcommand{\Pone}{\hbox{$P_{\rm 1}$}}
\newcommand{\Ptwo}{\hbox{$P_{\rm 2}$}}
\newcommand{\Ptwom}{\hbox{$P^{\rm m}_{\rm 2}$}}
\newcommand{\Ptwot}{\hbox{$P^{\rm t}_{\rm 2}$}}
\newcommand{\Pthree}{\hbox{$P_{\rm 3}$}}
\newcommand{\Pthreem}{\hbox{$P^{\rm m}_{\rm 3}$}}
\newcommand{\Pthreet}{\hbox{$P^{\rm t}_{\rm 3}$}}
\newcommand{\Pfour}{\hbox{$P_{\rm 4}$}}
\newcommand{\Fthree}{\hbox{$f_{\rm 3}$}}
\newcommand{\Fthreem}{\hbox{$f^{\rm m}_{\rm 3}$}}
\newcommand{\Fthreet}{\hbox{$f^{\rm t}_{\rm 3}$}}
\newcommand{\fN}{\hbox{$f_{\rm N}$}}
\newcommand{\D}{\hbox{$D$}}
\newcommand{\Dm}{\hbox{$D^{\rm m}$}}
\newcommand{\Dp}{\hbox{$D_{\rm p}$}}
\newcommand{\Dpt}{\hbox{$D^{\rm t}_{\rm p}$}}
\newcommand{\Nsp}{\hbox{$N_{\rm sp}$}}
\newcommand{\DelSig}{\hbox{$\Delta \sigma$}}
\newcommand{\DelEta}{\hbox{$\Delta \eta$}}
\newcommand{\DelPhi}{\hbox{$\phi_{off}$}}
\newcommand{\OurPsr}{\hbox{$\rm PSR~B0826-34$}} 
\newcommand{\ExB}{\hbox{${\rm {\bf E}\times{\bf B}}$}}

\begin{document}
\title{Unraveling the drift behaviour of the remarkable pulsar PSR~B0826$-$34}

\author{Y. Gupta\inst{1}, J. Gil\inst{2}, J. Kijak\inst{2} \and M. Sendyk\inst{2}}

\institute{National Centre for Radio Astrophysics, TIFR, Pune University Campus, Pune 411007, India
\and 
Institute of Astronomy, University of Zielona Gora, Lubuska 2, 65-265, Zielona Gora, Poland}

\offprints{Y. Gupta (E-mail: ygupta@ncra.tifr.res.in)}


\titlerunning{Unraveling drift behaviour of \OurPsr}
\authorrunning{Gupta et al.}

\abstract{
Pulsars with drifting subpulses are thought to be an important key
to unlocking the mystery of how radio pulsars work.  We present
new results from high sensitivity GMRT observations of \OurPsr\/ 
-- a wide profile pulsar that exhibits an interesting but complicated 
drifting pattern.  We provide a model to explain the observed subpulse
drift properties of this pulsar, including the apparent reversals
of the drift direction. In this model, \OurPsr\/ is close to
being an aligned rotator.  Using information about the polarization 
and frequency evolution of the pulse profile, we solve for the 
emission geometry of this pulsar and show that the angle between 
the rotation axis and the dipole magnetic axis is less than 5\degr.  
As a result, our line of sight samples a circular path that is 
entirely within the emission beam. We see evidence for as many as 
6 to 7 drifting bands in the main pulse at 318 MHz, which are all 
part of a circulating system of about 15 spark-associated subpulse 
emission beams that form, upon averaging, one conal ring of the mean 
emission.  We also see evidence for a second ring of emission, which 
becomes dominant at higher frequencies (above 1 GHz) due to the 
nature of the emission geometry. We model the subpulse drift behaviour
of this pulsar in detail, providing quantitative treatments of the 
aliasing problem and various effects of geometry which play an 
important role.  The observed drift rate is an aliased version of the 
true drift rate which is such that a subpulse drifts to the location 
of the adjacent subpulse (or a multiple thereof) in about one pulsar 
period. We show that small variations, of the order of 3-8\%, in the 
mean drift rate are then enough to explain the apparent reversals of 
drift direction seen in the data. We find the mean circulation time 
of the drift pattern to be significantly longer than the predictions 
of the original Ruderman and Sutherland (1975) model and propose 
an explanation for this, which relates to modified models with 
temperature regulated partial ion flow in the polar vacuum gap.  
The small variations in drift rate are then explained by very small 
heating and cooling effects -- less than 3500 K change in the 
$\sim 2.5\times10^6$ K surface temperature of the neutron star polar 
cap.  From a detailed consideration of the variation of the mean 
subpulse separation across the main pulse window, we show that the 
circulating spark pattern is {\it not}  centred around the dipole 
axis, but around a point much closer (within a degree or so) to the 
rotation axis. This is an indicator of the presence of a ``local pole'' 
corresponding to the non-dipolar magnetic fields that are expected to 
be present close to the neutron star surface.  \OurPsr\/ thus provides 
a very rich and powerful system in which to explore important aspects 
of the physics of pulsar radio emission and neutron star magnetospheres.
}
\maketitle

\section{Introduction } 		\label{sec:intro}

Understanding how radio pulsars work remains a challenging
unsolved puzzle in astrophysics. The phenomenon of drifting
subpulses seen in some pulsars is widely thought to be an
important key to understanding this mystery.  This phenomenon,
first reported by \cite{DC68}, manifests itself in the most basic
form as an organised phase change of subpulses occurring in
successive pulses, resulting in the formation of one or more drift
bands spanning the pulse envelope \citep{Backer73,JMR86}.
The basic concept is illustrated in figure 8 of \cite {Backer73}.
In any single pulse, there are typically two to three approximately
equispaced subpulses, with a separation of \Ptwo\/ degrees of
longitude, whose typical values are $\sim 10\degr-20\degr$.  The
interval between recurrence of successive drift bands at a given
pulse longitude is called \Pthree\/ and typical values range from 1
to 15 times \Pone, the pulsar period. The intensity of the
subpulses is systematically modulated as they move along the drift
band.  Usually, the intensity peaks in the middle of the pulse
envelope and decreases towards the edges; however, in a few cases,
the opposite trend is also seen \citep{TMH75}.
The direction in which the subpulses drift -- either leading to
trailing edge of the pulse window with pulse number, or the other
way around -- is another important aspect.  Though both kinds of
drifting pulsars are known, for a given pulsar, the direction in
which the subpulses drift is usually fixed. However, there are a
few exceptional pulsars where both directions of drift are seen
(though at different times).

In the standard phenomenology \citep[e.g.][]{JMR86}, cases of clear 
subpulse driftband patterns are usually seen in pulsars where the 
line of sight grazes along the edge of the radio emission beam.  
For pulsars where the line of sight has a more central traverse over 
the beam, the phenomenon can manifest itself as periodic intensity 
modulations of the subpulses at a given longitude, without the 
detection of any phase changes \citep{Backer73}.  The observed 
periodicities related to patterns of drifting subpulses are usually 
found to be independent of radio frequency, thus excluding all 
frequency dependent plasma effects as plausible mechanisms of the 
drifting subpulse phenomenon. All the properties discussed above 
strongly suggest that this phenomenon is a manifestation of the 
circulation of subpulse-associated subbeams around the magnetic 
axis.

The drifting subpulse phenomenon finds a natural explanation in
the model of \citet[][RS75 hereafter]{RS75}. This model proposes
the occurrence of discrete, localised spark discharges in a
quasi-steady vacuum gap that forms just above the neutron star
surface, as the seed activity that drives the radio emission
mechanism.  Each spark generates a stream of electron-positron
plasma that flows out along the dipolar field lines and, at
heights that are estimated to be of the order of several tens of
stellar radii, produces a beam of subpulse-associated radio
emission due to some, as yet unidentified, plasma radiation
mechanism. Further, this model proposes that the plasma of each
spark discharge inside the vacuum gap undergoes a systematic
motion around the magnetic axis, due to the \ExB\/ drift.  This,
coupled with the rapidly fluctuating but rarely self-quenching
nature of the sparks, leads naturally to the phenomenon of
drifting subpulses. This drift rate is slower than the corotation
rate and the sparks therefore lag behind with every rotation.  In
this picture, each drifting spark would correspond to one band of
drifting subpulses. Since the spark plasma can only lag the pulsar 
rotation, the model predicts drift in one direction only, depending 
on the relative orientation of the rotation and magnetic axes and 
the observer's line of sight for the given pulsar.  
The RS75 model also makes quantitative predictions of the total
circulation time of the drifting pattern for a pulsar, which, in
principle, can be checked against observations. Further
refinements of the RS75 model include the models of Cheng \&
Ruderman (1980), Gil \& Sendyk (2000, hereafter GS00) and 
Gil, Melikidze \& Geppert (2003, hereafter GMG03) which attempt 
to overcome some of the deficiencies and problems identified 
with the RS75 model.  Other models that address the issue of
subpulse drift are those of \cite{am83,j84,ketal96,w03}. 

However, there are some problems in the interpretation of the 
observations that make it difficult to compare with theoretical 
models.  Often, there are complications of the drift rates not 
being constant and stable with time.  Even when the drift rates 
(or value of \Pthree) are stable and can be determined reliably, 
it is not clear whether they are the real values or not, as the 
measured values can be modified due to the effect of aliasing. 
This effect can occur if the sampling rate (once per pulsar 
period) is too slow to track the same subpulse-associated subbeam 
of radiation in a sequence of pulses.  Further, the total number 
of sparks, \Nsp, needs to be known to ascertain the total 
circulation time of the pattern. In all this, the exact geometry 
of the observations needs to be known for correct interpretation 
of the results.

As a result, though there are about 40 pulsars with observations 
and measurements of the phenomenon of drifting subpulses 
\citep[for a summary see][]{JMR86}, there are only a few for which 
quantitative results have been obtained which can be compared 
against the predictions of the models.  For PSR B0943+10, from 
a careful analysis of the fluctuation spectra, \citet{DR99} were 
able to resolve the alias order for the observed periodicity 
(\Pthree) and also infer the presence of 20 spark associated 
subpulse columns circulating with a time of $37 \Pone~$.
For PSR B0809+74, \citet{vL_etal_2003} have recently shown 
that the drift rate we see is an unaliased version and hence 
$\Pthree\/ > 150\Pone~$ and $\Nsp\/ \ge 14$. For both cases, the 
final circulation time turns out to be significantly longer than 
the prediction of the RS75 model. Modified versions of this model 
(e.g. GMG03) attempt to account for these discrepancies.

Whereas observations of pulsars with drifting subpulses where the
line of sight just grazes a cone of emission produces a clear
signal for analysis and interpretation, a similar phenomenon when
seen for pulsars where the line of sight samples a large fraction
of the conal ring can produce additional insight into the phenomenon.  
This makes pulsars with wide profiles very interesting.  \OurPsr\/ 
is such a case. It has a very wide profile, showing emission over 
as much as 250\degr\/ of longitude at metre wavelengths 
\citep{Durdin,Biggs,Reyes}.  However, it is not a widely studied 
pulsar, probably because of the fact that it nulls for as much as 
70\% of time \citep{Durdin}, making it a difficult case to observe. 
The study reported by \citet{Biggs} is the most exhaustive study 
published to date for this pulsar.  They found that there is 
subpulse structure at almost all longitudes in the wide profile of 
\OurPsr, with remarkable patterns of drifting subpulses present 
over 200\degr\/ of longitudes.  They noted that the drift rate 
shows large variations, including reversals of the direction of the 
drift -- they concluded that this behaviour is inconsistent with the
prevailing theoretical models for the phenomenon of drifting
subpulses.  From the integrated pulse profiles and the polarization 
data at 408 and 610 MHz, \citet{Biggs} concluded that \OurPsr\/ is 
likely to be a nearly aligned rotator, i.e. with rotation and magnetic 
axes almost parallel to each other, although their formal best fit
values for the angle between the rotation and magnetic axes don't
fully support this conclusion. However, in the morphological
scheme of \citet{JMR83,JMR93} it is classified as ``M'' type
pulsar, i.e. having a central line of sight which goes close to
the magnetic pole and produces distinct core and conal emission
components in the average profile.

It is clear from the above description that \OurPsr\/ is an
enigmatic pulsar that has the potential to reveal vital clues
about the pulsar emission mechanism and therefore deserves a
closer and more detailed study.  With this in aim, we have 
been carrying out high sensitivity observations of this pulsar with 
the Giant Metre-wave Radio Telescope (GMRT).  The results from our 
initial observations and the interpretations thereof are reported 
in this paper. In section \ref{sec:obs}, we describe the observations 
and the basic data reduction.  Section \ref{sec:RnI_main} presents 
the main results and our model that interprets these results, 
including determination of the emission geometry for this pulsar 
and detailed description of various effects that modify the true
values of basic quantities related to the drifting phenomenon. In
section \ref{sec:discuss} we discuss the implications of different
aspects of our model and present a final summary. In Appendix A we
summarise the basic ideas relevant to the understanding of pulsar
emission geometry, and in Appendix B we describe the details of
our simulation studies.

\section{Observations and Data Analysis}    \label{sec:obs}

PSR~B0826$-$34 was observed on Dec 11, 2000 at the (GMRT), in 
the 325 MHz band, using the incoherent array mode \citep[see]
[ for more details about the pulsar modes of operation of the GMRT]
{Gupta}. The data were obtained by incoherent addition of the dual 
polarization signals from 13 antennas.  The center frequency of the 
observations was 318 MHz. The bandwidth used was 16~MHz, divided 
into 256 spectral channels by the digital back-ends.  The raw data 
were integrated to a time resolution of 0.516 milliseconds before 
being recorded for off-line analysis, where they were further 
integrated to obtain a final time resolution of 4.128 milliseconds.  
The duration of the observation was about 22 minutes, resulting in
recorded data covering 714 periods of this pulsar. Of this, only
the first 500 or so pulses were useful; the pulsar was in the null
state after that, and no detectable signal could be obtained from
the last 200 pulses.

During off-line analysis, the data were first dedispersed and then
folded synchronously with the {Doppler corrected} pulsar period to
obtain the average profile, which is shown in the second panel of
Figure \ref{fig:freq_evln}.  The first remarkable thing about the
average profile for \OurPsr\/ is the unusually large pulse width :
emission can be seen over more than 200\degr\/ of longitude.  The
``main pulse'' has a double peaked structure with a component peak
separation of $\rm 134\degr\/ \pm 4\degr$, centred on the
longitude of 240\degr.  In addition, there is evidence for faint
emission in the longitude range 55\degr to 105\degr, which could be 
an interpulse, centred about 90\degr.  These features are similar to
what is seen in the 408 MHz profiles of \citet{Biggs} and
\citet{Durdin}.

Comparing all the published average profiles available for this pulsar
at different frequencies \citep[][ the EPN database]{Biggs,Durdin,Reyes},
it is clear that there is a remarkable evolution of the pulse profile
with frequency.  It appears that the main pulse reduces in strength relative
to the interpulse systematically, as one goes from lower to higher
frequencies, to the extent that the interpulse actually dominates over
the main pulse at frequencies above $\sim$ 1 GHz (see lower 3 panels of
figure \ref{fig:freq_evln}, for example).  Further, the component
separation between the two peaks of the main pulse reduces with increasing
frequency, following the general trend seen for a large majority of pulsars
(referred to as the ``radius-to-frequency mapping'' model).
In Table {\ref{tbl1}, we quantify this frequency evolution by compiling
all the reported values for the component separation for the main pulse
available for this pulsar, including our new results which are found to
be compatible with the trend.  The separation between the centres of the
main pulse and the interpulse appears to be roughly frequency independent,
at a value of about 150\degr\/ to 160\degr\/ (see figure \ref{fig:freq_evln}).

As reported by \citet{Biggs}, \OurPsr\/ exhibits remarkable
subpulse structure in single pulse data.  Figure
\ref{fig:stack_line} shows an example of single pulse data,
covering the first 200 pulses from our observations.  Individual
subpulses can be seen over almost the entire main pulse window.
Further, there is evidence for regions of organised drifting of
the subpulses (in the form of linear drift-bands going from later
to earlier longitudes with increasing pulse number -- we refer to
this as negative drift) over limited period ranges.  There is also
evidence for regions of oppositely oriented drift-bands (positive
drift rate), as well as clear signs of curved drift-bands.  In
addition, there are sporadic, short duration nulls interspersed in
the data.  The over-all picture thus looks somewhat random and
chaotic, and not very different qualitatively from what was reported
by \citet{Biggs} for this pulsar at 645 MHz (their Fig.~2).

However, the better signal-to-noise ratio of our data allows a 
clearer picture of the drifting subpulse pattern to emerge, from a
gray-scale plot of the single pulse data (Figure \ref{fig:stack_grey}).
Here, one can clearly see evidence for multiple drift-bands across 
the pulse window; 6-7 bands can be seen in most of the sections 
where the pulsar does not null.  However, the drift behaviour of 
these bands of subpulses is not a regular, monotonic, linear drift 
pattern; instead it is much more irregular, exhibiting significant 
changes in the apparent drift rate, including change of sign. 
Nevertheless, the pulse regions showing drift rates of opposite sign 
are almost always connected by regions showing a smooth transition of 
the drift rate.  This can clearly be seen for most cases where the 
apparent drift rate changes from negative to positive values, going 
smoothly through zero (e.g. see the regions of pulse \# 20 to 70, pulse 
\# 120 to 180 and pulse \# 350 to 390).  For the opposite transition 
(from positive to negative drift values), the change is usually more 
abrupt and not always easily traceable.  Such regions (e.g. pulse 
\# 80 to 120 and pulse \# 180 to 220) correspond to the ``run-away'' 
sections identified by \citet{Biggs}. The region of pulse \# 390 to 440 
is an exception, where the complete transition can be traced. Therefore, 
in the most general case, the drift pattern for this pulsar consists 
of connected regions of drift-bands of opposite slopes. The typical
values for the maximum positive and negative drift rates are
+1\fdg9/pulse (about +1\fdg0/sec) to -0\fdg8/pulse (about
-0\fdg4/sec). Further, the different curved sections of drift
bands can be connected with a continuous curve, except for missing
sections due to nulls. A typical time scale for one cycle of drift
rate reversal is about 100 periods, and about 5 such cycles are
present in the sequence of 500 pulses in Figure
\ref{fig:stack_grey}. It is also very significant that the
behaviour of all the visible drift bands is highly correlated,
i.e. all of them simultaneously show the same kinds of changes in
drift rate.

Since there are about 7 drift bands present in the main pulse
region of about 160 degrees, the typical separation between drift
bands works out to about 23 degrees.  A quantitative estimate of
the separation between adjacent drift-bands -- traditionally
referred to as \Ptwo\/ -- is obtained from the autocorrelation
function of the single pulses, computed over the entire pulse
window and averaged over all the available pulses. The result is
shown in Figure \ref{fig:full_acf}, where the secondary peak is
due to the correlation between adjacent drift-bands.  From the
location of this peak, the average value of \Ptwo\/ is found to be
$\rm24.9 \pm 0.8$ degrees, close to the rough estimate given above.
This is somewhat smaller than the value of $\rm29.0 \pm 2.0$ obtained
by \citet{Biggs} at 645 MHz.  In addition, there is some indication
from our raw data itself (e.g.  Figure \ref{fig:stack_grey}), of
variation of the value of \Ptwo\/ across the pulse window, with a
tendency to increase from the centre to the edge of the main pulse.
We return to this aspect in detail in section \ref{sec:interpret_2}.

\section{Results and Interpretation}  \label{sec:RnI_main}

To summarise the main results from our analysis : \OurPsr\/ has a
unusually wide average profile and the single pulse data show
multiple drifting bands of subpulses which drift in a highly
correlated manner but show significant changes in the apparent
drift rate, including reversals of the drift direction. Our basic
model to explain this behaviour is as follows:  This pulsar is an
almost aligned rotator (i.e. dipole magnetic axis and rotation
axis are almost aligned).  As a result, our line of sight traces
out a circular track through the emission beam, centred very close 
to the magnetic axis.  This allows us to sample the radiation from 
a large number of subpulse-associated plasma columns present in one 
particular conal ring of the emission beam, resulting in the large 
number of drifting bands of subpulses that are seen. The 
subpulse-associated plasma columns are produced by an arrangement 
of spark discharges occurring in the vacuum gap above the polar cap, 
as envisaged in models such as in RS75 and GS00.  The number of 
such sparks that need to be present in this particular ring, to 
explain the observed number of drift bands, is about 15. These 
spark discharges circulate in closed tracks on the polar cap, 
presumably due to the \ExB\/ drift.
Further, the observed drift rate of the these bands is not
the intrinsic drift rate, but an aliased version.  In this case,
the intrinsic drift rate is such that during one rotation period,
a subpulse drifts to reach the location of the adjacent subpulse,
or a multiple thereof; i.e. \Pthree\/ -- traditionally defined as
the time interval required for the subpulse pattern to repeat at
the same longitudes -- has a value $\approx \Pone/k$ (where k is
an integer and \Pone\/ is the pulsar period).  If the drift rate
is such that $\Pthree\/ = \Pone/k$ exactly, then the observed
drift-bands would be longitudinally stationary, i.e. zero apparent
drift rate.  Now in addition, if there are small fluctuations in
the intrinsic drift rate around this mean value, then the slowing
down and speeding up of the intrinsic drift rate would be seen as
apparent drifts in opposite directions.  Such stroboscopic effects
would then give rise to the apparent curved drift-bands that are
observed in this pulsar.  The small fluctuations of the drift rate
could be produced by thermal effects on the polar cap that
regulate the ion flow in the acceleration gap \citep[][GMG03]{CR80}. 

We now discuss different aspects of our model in
detail.  We start with a determination of the emission geometry
for this pulsar, using information about its polarization
properties and frequency evolution of the profile (section
\ref{sec:geometry}). This is followed by a detailed modeling of
the drift behaviour (section \ref{sec:drift}), including the
effect of aliasing and the effect of geometry on the estimation of
\Ptwo.  In the light of these, we present our final interpretation
of the data (section \ref{sec:interpret_2}), deriving the possible
values for the various drift related parameters for this pulsar.

\subsection{Determination of the geometry}  \label{sec:geometry}
Determination of the emission geometry is vital for proper interpretation of the
results.  The most important parameters here are the following: \A, the angle
between the rotation and magnetic axes; \B, the angle of closest approach between
the magnetic axis and the line of sight (also called the impact parameter); and
$\rho$, the radius of the emission beam which, in general, is a frequency dependent
quantity.  The beam radius can also be replaced by specifying the parameters $s$
(normalised distance from the dipole axis to the foot of the dipolar field line,
a frequency independent quantity) and $r_{em}$ (the emission height of the radio
radiation, a frequency dependent quantity).  The basic equations describing the
relationships between these quantities are recapitulated in Appendix~A.

Traditionally, the values of \A\/ and \B\/ for a pulsar are
determined from fits to the polarization angle (P.A.) curve of the
pulsar, using the rotating vector model (eqn.\ref{eqn:PA}).
However, as has been shown by several authors \citep {Miller93,
vOmmen97}, there are problems in obtaining reliable, independent
estimates of these angles from this procedure, in that a wide
range of combination of values of \A\/ and \B\/ are found to give
acceptable fits. For the specific case of \OurPsr, \citet{Biggs}
have shown that the polarization angle swing is compatible with a
large range of choices of \A\/ and \B. Furthermore, this method
does not provide any information about the beam radius.  The
latter is usually estimated from a suitable measurement of the
pulse width (such as separation between component peaks, 10\%
width etc) and the use of eqn. \ref{eqn:gamma}, after the values
for \A\/ and \B\/ have been determined (see Appendix A).

To determine the emission geometry for \OurPsr, we have followed
a slightly different technique which combines the available
polarization information with the measured frequency evolution of
the component separation, to constrain the values of the geometry
parameters. Specifically, we use the following constraints: value
of the steepest gradient of the P.A. curve, which we estimate to
be $2.0 \pm 0.5$ degrees per degree of longitude, from figure 6 of
\citet{Biggs} (we note that \cite{LM88} quote $1\fdg7 /\degr\/$ 
for this value); the total swing of the polarization angle over the 
main pulse region, which we estimate to be $ \approx 90\degr\/$, 
again from figure 6 of \citet{Biggs}; the measured values of the 
separation of the main pulse components, as listed in our Table 
\ref{tbl1}.  We performed a grid search in values of \A, \B\/ and 
$s$ to find all possible combinations of values that satisfy these 
constraints at all the frequencies listed in Table \ref{tbl1} 
(relevant details are described in Appendix A).  Viable solutions 
for \A\/ and \B\/ were found for values of the $s$ parameter in the 
range $0.25 \le s \le 0.85$.  These solutions were in the range 
$1\fdg5 \le \A\/ \le 5\fdg0$ and $0\fdg6 \le \B\/ \le 2\fdg0$, 
with the value of $\A\/$ greater than $\B\/$ such that the requirement 
of equation (\ref{eqn:gradient}) was satisfied.  The extremes of the
solution range were : $s=0.25,\ \A\/=1\fdg5,\ \B\/=0\fdg6$ for the 
smallest sized conal ring to $s=0.85,\ \A\/=5\fdg0,\ \B\/=2\fdg0$ 
for the largest sized conal ring.  Further, from the typical S-shaped 
P.A. curve \citep[] [and also the upper panel in our figure
\ref{fig:freq_evln}] {Biggs}, it is clear that we have a case of
an ``outer'' line of sight, i.e. a case where the magnetic axis
lies between the rotation axis and the line of sight, giving a
positive value for \B\ \citep[see][]{nv82}.

It is indeed remarkable that we are able to obtain such good fits
matching the profile evolution over the frequency range of 82 MHz
to 610 MHz, while meeting the basic polarization angle requirements. 
Based on fits to polarization data only, \citet{Biggs} obtained very 
different values for \A\/ and \B\/ for this pulsar.  Using their 
values, it is impossible to reproduce the observed pulse widths for 
any realistic emission height model.  In the allowed range of 
$0.25 \le s \le 0.85$, we tend to favour smaller values of $s$ 
($\approx$ 0.25 to 0.5), as we believe that there is another, outer 
conal ring of emission for this pulsar, with an $s$ value of 
$\approx$ 0.7 (see section \ref{sec:discuss}).  As a typical set 
of values, we use $s=0.44,\ \alpha=2\fdg5,\ \beta=1\fdg0$ 
wherever specific values are required in the rest of this paper. 
For this specific set of values of $s,~\alpha$ and $\beta$, the 
predicted separation between the component peaks is 
$116\degr,\ 124\degr,\ 130\degr,\ 152\degr,\ 158\degr$ and
$159\degr$ at frequencies of 610, 408, 318, 116, 95 and 82~MHz,
respectively, which agree very well with the measurements listed 
in Table \ref{tbl1}.
We note that \citet{LM88} report best fit values of $\A=2\fdg1,
\ \B=1\fdg2$ (their Table~1) and \citet{JMR93} reports $\A=3\degr,
\ \B=1\fdg1$, both of which are compatible with our results. 

These results confirm our first hypothesis that we are looking at
a pulsar that is very close to being an aligned rotator. 

\subsection{Understanding the drift behaviour}  \label{sec:drift}
The basic observable quantities involved in understanding the subpulse
drift pattern for a pulsar are illustrated in figure 8 of \citet{Backer73}.
We start our discussion with the idealized case of a perfectly aligned 
rotator ($\alpha = 0\degr$), in which case the line of sight traces out 
a circle on the polar cap, centred exactly around the magnetic axis. 
Further, we assume that the line of sight is exactly coincident
with a conal ring of emission.  In such a case, regularly spaced
subpulses would be visible throughout the pulsar period. Later we
modify our discussion to include realistic cases.
The principal periodicities that can be measured are \Pone, \Ptwo,
\Pthree. The \Pthree\/ periodicity is often expressed in terms of a
``fluctuation frequency'', \Fthree\/ $\equiv$ 1/\Pthree. These can be
related to the drift rate, \D, as follows :
\begin{equation}
\Fthree ~~=~~ \D / \Ptwo ~=~ \Dp / (\Pone\/ \Ptwo) ~~~,       \label{eqn:drift1}
\end{equation}
where \Pone\/ and \Pthree\/ are in units of seconds (\Fthree\/ is in Hz), \Ptwo\/ is
in degrees, \D\/ is in degrees per second and \Dp\/ is in degrees per period.
If there are \Nsp\/ equispaced sparks in a circular ring on the polar cap, then
\begin{equation}
\Ptwo ~~=~~ \frac {360\degr}{\Nsp} ~~~.    \label{eqn:P2-Nsp}
\end{equation}
Further, the total circulation time of the pattern, i.e. the time it would
take a drifting spark to complete one full circulation, can be expressed as
\begin{equation}
\Pfour ~~=~~ \Nsp \Pthree ~~=~~ \frac {360\degr} {\D} ~~~,
\label{eqn:P4-Nsp}
\end{equation}
where \Pfour\/ is measured in seconds.
Thus, in principle, the measured values of \Pone, \Ptwo\/ and \Pthree\/ can be used to
find the drift rate and the total circulation time of the spark pattern, which can
be compared with the predictions of theoretical models.

In practice, however, the situation is complicated by the fact that the measured
values of \Ptwo\/ and \Pthree\/ (\Ptwom\/ and \Pthreem) need not correspond to the true
values (\Ptwot\/ and \Pthreet); \Ptwo\/ can be affected by the drift rate and by the
viewing geometry, and \Pthree\/ can be affected due to aliasing.  Also, some of the
above equations are not valid for the general case of the non-aligned rotator.
We first consider the effect of aliasing on \Pthree\/ and then examine the factors
affecting \Ptwo\/ and the effect of the viewing geometry.

\subsubsection{Effect of aliasing on drift behaviour}   \label{sec:aliasing}
Since the signal at any given pulse longitude is sampled only once
per pulsar rotation period, the interpretation of any periodic
signal measured at that longitude (such as the periodicity
represented by \Pthree) is subject to the constraints of this
sampling rate.  For sufficiently slow drift rates (such that
$\Pthreet\/ > 2\Pone$, or $\Fthreet\/ < 0.5/\Pone$), \Pthreem\/ is
the same as \Pthreet.  For $\Pthreet\/ = 2\Pone$, a subpulse
drifts to the longitude of the adjacent subpulse in exactly 2
periods. This corresponds to the Nyquist periodicity, with $\fN\/
= 0.5/\Pone\/$ being the Nyquist frequency.  For faster drift
rates (such that $\Pthreet\/ < 2\Pone$, or $\Fthreet\/ > \fN$), an
aliased version of the periodicity is seen.  For \Pthreet\/
slightly less than $2\Pone$, \Pthreem\/ is slightly more than
$2\Pone\/$ and the apparent drift is of opposite sign than the
true drift.  As \Pthreet\/ reduces further, \Pthreem\/ keeps
increasing correspondingly, till $\Pthreet\/ = \Pone$, at which
point \Pthreem\/ becomes infinite ($\Fthreet\/ = 2 \fN ~,~~
\Fthreem\/ = 0$) -- this corresponds to the case of longitude
stationary subpulse patterns, with zero apparent drift.  As
\Pthreet\/ reduces further ($\Fthreet\/ \ge 2 \fN$), \Fthreem\/
begins to increase and a non-zero drift rate is again measurable;
however, the direction of the drift is now the same as the true
direction. In fact, the apparent drift direction reverses sign
every time the true drift rate crosses a multiple of the Nyquist
boundary ($n~\fN$), whereas the apparent drift amplitude goes
through zero for every ``even'' crossing (n=even number) and
through the maximum observable value for every ``odd'' crossing
(n=odd number). In Appendix~B we present results from simulations 
which illustrate some of these effects.

Quantitatively, the above behaviour can be described as follows :
\begin{equation}
\Fthreet ~~=~~ 2 k \fN ~+~ (-1)^l \Fthreem ~~,~~~~  \Fthreem ~~=~~ \frac {\Dm}{\Ptwom} ~~~;   \label{eqn:alias-drift}
\end{equation}
where $k = INT[(n+1)/2]$ and $l = mod(n,2)$, and $n$ is the alias order such that
$n \fN\/ < \Fthreet\/ < (n+1) \fN$.  This can be considered as a generalised form of
eqn. \ref{eqn:drift1}, which special case obtains for $n=0$.

Thus, in order to estimate the true value of \Pthree\/ (and hence determine \Nsp\/ and 
\Pfour\/ in accordance with eqn. \ref{eqn:P4-Nsp}) the alias order, $n$, has to be 
resolved; only for $n=0$ is there a one to one correspondence between the measured 
and true quantities.  In general, it is not always possible to infer the value of 
$n$ from the observations.  As described in section \ref{sec:intro}, it has been 
achieved for only 2 pulsars -- PSR B0943+10 \citep {DR99} and PSR B0809+74 
\citep {vL_etal_2003} -- using extra information from specific features present 
in the data.

For \OurPsr, our proposal is that we are seeing an aliased drift rate such
that $n=2,4,6\ldots$ (i.e. $k=1,2,3\ldots$) and the overall mean drift rate is such
that $\Fthreem ~\approx~0$.  Thus $\Fthreet\/ \approx 2k\fN$.  In addition, there are
small variations in $\Fthreet\/$ with an amplitude $\mid {\Fthreem} \mid \ll \fN$,
but large enough such that it crosses back and forth across the Nyquist boundary ($n$
goes to $n-1$, $k$ remains the same, $l$ changes sign), producing the observed pattern
with apparent reversals of drift direction.

The effect of aliasing also complicates the effort to resolve the
direction of the true drift and compare it with model predictions.
In vacuum gap models, the \ExB\/ drift of the spark plasma in the
gap is slower than the corotation speed, i.e. the sparks lag
behind with respect to the rotation of the star (e.g. RS75). The
direction of the measured subpulse drift that this translates to
depends on the line of sight: for an ``inner'' line of sight
(negative \B), the subpulses drift from the trailing edge to the
leading edge of the pulse window with increasing pulse number,
i.e. negative drift; for an ``outer'' line of sight (positive \B),
the drift is from leading to trailing (positive drift). However,
for a case where \A\/ is very small, such that the spin axis lies
within the emission cone ($\A\/ \le \rho$), there is only one
direction of drift for any kind of line of sight: drift from
leading to trailing (positive drift) -- see \citet{R76} for
similar arguments.  Of course, the final observed direction of the
drift is further modified by the aliasing effect as described
earlier in this section.

For \OurPsr, we have shown that we are dealing with an almost
aligned rotator, with values of \A\/ that are smaller than the
angular size of the ring of sparks (section \ref{sec:geometry}).
Further, we are quite certain that we are dealing with an
``outer'' line of sight geometry.  Both these factors imply that 
the intrinsic drift rate should be positive (i.e. drift from 
leading to trailing in the pulse window).  For an unaliased drift, 
the direction of the apparent drift will be same as that of the 
intrinsic drift. As shown earlier, the apparent drift reverses 
direction each time the amplitude of the intrinsic drift rate 
crosses a multiple of the Nyquist boundary.  Now, in our model 
the intrinsic drift rate for \OurPsr\/ is such that we are close 
to an even Nyquist\footnote{In fact, for odd Nyquist boundaries
(n=1,3,5,...) one would observe clear ``even-odd'' intensity
modulations like in PSR B0943+10 (Deshpande \& Rankin 1999, GS00
[their Fig.~2]) and PSR B2303+30 (GS00 [their Fig.~3]),
which are, however, not observed in PSR B0826-34 (see Appendix B
for further discussion of aliasing effects).} boundary
($n=2,4,6...$), with small variations around this value. When the
intrinsic drift rate is slightly above the Nyquist boundary, the
apparent drift direction should be same as the intrinsic.  This
means that the regions of positive drift in figure
\ref{fig:stack_grey} are regions where the drift rate is slightly
faster than the mean rate and regions of negative drift are
regions where the drift pattern slows down somewhat.  Coupling
this with the drift behaviour described in section \ref{sec:obs},
we conclude that the transition when the drift rate speeds up from
less than the mean value to over the mean value is a relatively
smoother and slower process than the opposite change when the
drift rate slows down, which is more abrupt and faster.  The
implications of this are discussed further in section~\ref
{sec:discuss}.

\subsubsection{Effect of geometry on \Ptwo} \label{sec:P2}
The measured value \Ptwom\/ does not correspond to the true value \Ptwot, due to 
two effects.  The first is the effect of the finite time taken by the line of sight 
to traverse the beam.  As the line of sight moves from one subpulse longitude to 
the next one, the subpulse pattern drifts in longitude by a small amount; as a
consequence, the measured value \Ptwom\/ is a ``Doppler-shifted'' version of 
the actual subpulse separation.  This is illustrated schematically in Figure 
\ref{fig:P2_stretch}.  It is straightforward to show that \Ptwot\/ and \Ptwom\/ 
are related by
\begin{equation}
\Ptwot ~~=~~ \Ptwom(1 ~-~ \Dpt/360^{\circ}) ~~~,            \label{eqn:P2-stretch}
\end{equation}
where, in keeping with our convention, the sign of \Dpt\/ is positive for drift from
leading to trailing longitudes (i.e. along the direction of traverse of the line of
sight), and negative for the other direction.  This modified value of \Ptwo\/ needs to
be used in equations \ref{eqn:drift1} and \ref{eqn:P2-Nsp} above.
In our model for \OurPsr, taking the slowest drift rate (corresponding to the alias
order of $n=2$ or $k=1$) and the mean value of \Ptwom\/ found in section \ref{sec:obs},
this correction amounts to $\approx 7\%$, and increases by the same amount for every
increment in the value of $k$.

Finally, for the general case of the non-aligned rotator, the
direct relation between \Ptwo\/ and the angular spacing of the
sparks on the polar cap (eqn. \ref{eqn:P2-Nsp}) is no longer
valid.  This is because equal longitude intervals along the line
of sight (which are measured with respect to the rotation axis) do
not map to equal angular intervals (magnetic azimuth) of the
corresponding events on the polar cap ring (which are measured
with respect to the magnetic axis).  The general formula that
relates azimuthal angles measured with respect to the magnetic
axis $(\sigma)$ with the rotational phase $(\phi)$ is
\begin{equation}
sin(\sigma) ~~=~~ \frac {sin(\alpha + \beta) sin (\phi)} {sin(\Gamma)} ~~~,    \label{eqn:sigma-phi}
\end{equation}
where $\Gamma$, the angle between the magnetic axis and the line of sight at
the rotational phase $\phi$, is determined from equation \ref{eqn:gamma}.
Figure \ref{fig:sigma_plot} shows the variation of $\sigma$ as a function of $\phi$
for a couple of sample cases of emission geometry that are relevant for \OurPsr.

As a consequence of this effect, equation \ref{eqn:P2-Nsp} can not be used as it
stands to compute \Nsp.  Instead, from a measurement of the longitudes of the
subpulses, $\phi_{i}$ (or of the intervals between them, $\Delta\phi_{i}$ = \Ptwom),
one first needs to correct for the effect of scaling due to drift
(eqn. \ref{eqn:P2-stretch}) and then map these corrected angles (or the intervals
between them) to corresponding angular values ($\sigma_{i}$ or $\Delta\sigma_{i}$)
on the polar cap ring.  For the correct solution, the values $\sigma_{i}$ will be
equispaced, and this constant interval will be
\begin{equation}
\DelSig ~~=~~ \frac {360\degr}{\Nsp} ~~~, \label{eqn:del-sigma}
\end{equation}
provided that the subpulse-associated subbeams circulate
around the magnetic dipole axis.

The conclusions from this section on understanding the drift behaviour can be
summarised as follows.  In order to estimate the total circulation time (\Pfour),
values of \Pthreet\/ and \Nsp\/ need to be estimated (eqn. \ref{eqn:P4-Nsp}).  The
former requires knowledge of the alias order (eqn. \ref{eqn:alias-drift}), and
the latter requires knowledge of \Ptwot\/ (eqn. \ref{eqn:P2-Nsp}).  \Ptwot\/ can
be estimated from the variation of \Ptwom\/ (eqns. \ref{eqn:P2-stretch},
\ref{eqn:sigma-phi} and \ref{eqn:gamma}), provided the geometrical parameters
($\A,\ \B$) are known and the drift rate is known, which in turn requires knowledge
of \Pthreet\/ and \Ptwot.
Clearly, these equations form a coupled set which can not be solved directly to 
obtain the values of $\Nsp,n$ (and hence of \Pthree\/ and \Pfour); instead, some 
kind of iterative or grid search method is needed to find the allowed solutions.

\subsection{Interpretation of the data}     \label{sec:interpret_2}
We now return to a closer inspection of our single pulse data to see what it can
tell us about the nature of the variation of \Ptwo\/ across the main pulse window.
The correlation analysis that was used to estimated the mean value of \Ptwom\/ (as
discussed in section \ref{sec:obs}) was repeated over several narrow windows
of pulse longitude (of about 40 degrees width), which were shifted systematically
across the entire main pulse window (in steps of about 4 degrees).  For each case,
the secondary peak of the correlation function was computed (where possible),
providing an estimate of the value of \Ptwom\/ in that longitude range.  Figure
\ref{fig:window_corr} shows sample plots of the correlation functions for 5 different
windows, exhibiting significantly different values of \Ptwom.
Figure \ref{fig:P2_varn} (top panel) shows the variation of \Ptwom\/ with pulse
longitude, compiled from this correlation analysis.  The missing data points
correspond to cases where the correlation function did not show a clear secondary
peak -- these are regions of the pulse window where the drifting subpulse signal
is relatively weak.  The error bars for the \Ptwom\/ estimates are not from any
rigorous calculations or curve fitting; they are estimates by eye, based on the
quality (level of noise and the shape of the curve around the secondary maxima)
of the cross correlation functions.

As seen in the top panel of figure \ref{fig:P2_varn}, there appears to be a smooth
trend for the variation of \Ptwom\/ across the pulse window, except for a few
longitudes where there is a sudden, sharp increase in the estimated value.  These
outliers (four are clearly identifiable in the figure) are not well understood and
are most likely due to the window spanning across some pairs of very strong subpulses
associated with some of the drift-bands; we ignore these four data points in the
subsequent analysis.  The figure shows that there is a measurable variation of 
\Ptwom\/ across the main pulse window, which can be seen more clearly in the trailing 
half of the profile.  The range of variation is from a minimum of about 21\fdg5 to 
a maximum of 27\degr, and is compatible with the mean value of 24\fdg9 that we
obtained, in section \ref{sec:obs}, from the correlation function for the entire
main pulse.  The other interesting thing is that the minimum is not centred
exactly in the middle of the on-pulse window, but appears at somewhat earlier
longitudes, by about 15\degr\/ or so (though the large error bars make it difficult
to come to very firm conclusions about this).

We now ask the question if our inferred geometry (as derived in section 
\ref{sec:geometry}) can explain this observed dependence of \Ptwom\/ 
with longitude.  The solid line in fig.  \ref{fig:sigma_plot} shows the 
variation of $\sigma$ with $\phi$ for the case of the 
$\A,\B\/ = 2\fdg5,1\fdg0$, which is in middle of the solution range 
found in section \ref{sec:geometry}.  It is clear from this plot that 
the variation of $\sigma$ is too fast for smaller values of $\phi$ and 
will lead to an over-correction of the smaller \Ptwom\/ values (in the 
middle of the pulse).  This is true even after the measured subpulse
longitudes are scaled to accommodate the effects of drift as described 
by eqn. \ref{eqn:P2-stretch}.  In fact, all the solutions found in section 
\ref{sec:geometry} yield geometries where the line of sight is too 
``shallow'' with respect to the emission ring, i.e. it is not sufficiently 
concentric with the emission ring.  This is primarily because the value of 
\A\/ is too large, relative to that of \B. For such geometries, not only 
is the observed variation of \Ptwom\/ across the pulse window incompatible 
with the predictions, the observed intensity levels in the saddle region 
of the main pulse are also much higher than the predictions, i.e. the central 
bands of drifting subpulses should be much weaker than what is observed. 
This is illustrated by the dotted-line simulated profile in Fig.~1 (see 
Appendix~B for a description of the simulation technique), which is
obviously inconsistent with the observed 318 MHz profile -- the components 
are too narrow and the saddle is too weak. 

To explain the apparent disagreement between the emission geometry
and the observed detailed behaviour of the subpulse pattern, we
propose the following refinement of our basic model.  Since the
emission geometry constraints from pulse width and polarization
data (section \ref{sec:geometry}) basically depend on the dipole
field geometry at significant heights (roughly in the range 10-50
stellar radii), whereas the detailed variations of the subpulse
structure are determined by events very close to the neutron star
surface (in the ``vacuum gap''), we propose that the centre of
activity of these two are associated with different locations in
the magnetosphere.  More specifically, we propose that the spark
pattern is not centred around the dipolar axis of the pulsar, but
instead, circulates around some other point on the polar cap.  In
models which propose higher order or ``multi-polar'' magnetic
field structure very near to the neutron star surface
\citep[e.g.][ and references therein]{GMM2002}, this alternate
point of circulation could be the location of the ``local magnetic
pole'' which, in general, could be offset from the dipole magnetic
pole. In such a scenario, the generation of the spark pattern in
the vacuum gap region is controlled by the local magnetic pole,
with concentric rings of sparks being formed and circulating
around this local pole.  The subpulse-associated plasma columns
generated by this activity then stream out into the magnetosphere
along the dipolar field lines, which dominate the magnetic field
structure above a certain height (certainly in the regions where
we believe the radiation to originate). The location of the local
pole (offset from the dipole) can be specified in terms of a
new set of angles \Ap\/ and \Bp\/ analogous to \A\/ and \B\/ (note
that $\Ap + \Bp = \A + \B$ needs to be satisfied), and a rotational
angle specifying the rotational phase difference $\phi_{off}$
between the fiducial plane containing the rotation axis and the
dipole axis and the fiducial plane containing the rotation axis
and the local pole axis. 

From the variation of \Ptwom\/ (figure \ref{fig:P2_varn}, top panel) 
and the model curves in figure \ref{fig:sigma_plot}, it is clear 
that in our model for \OurPsr, the local pole needs to be located
significantly closer to the rotation axis that the dipole axis.
Further, a rotational phase difference of about $15\degr\/$ is
suggested by the location of the minimum of the \Ptwom\/ curve in
the top panel of Fig.~\ref{fig:P2_varn}.  How can we make a more
quantitative determination of the local pole geometry? We need
simultaneous solutions for the following: values of \Ap, \Bp\/ and
\DelPhi\/ for the local pole, value of \Nsp\/ and the alias order
$k$, such that the inferred values of angular separation of the
sparks circulating around the local pole (we will use $\eta$ to
refer to these angles, to differentiate from $\sigma$ used for
angles with respect to the dipole axis), will satisfy the
corresponding version of equation \ref{eqn:del-sigma}. To achieve
this, a grid search in the values of \Nsp, $k$, \Ap\/ and \Bp\/
was carried out. For each possible solution point in the grid
search, a few different choices of \DelPhi\/ were tried. For each
combination, the \Ptwom\/ curve (top panel in fig \ref{fig:P2_varn}) 
was ``corrected'' to obtain a curve of \DelEta\/ versus longitude 
(lower panels of fig \ref{fig:P2_varn}) as follows. The measured 
\Ptwom\/ values at different phase locations ($\phi_i$) were 
converted to a pair of individual phase locations 
($\phi_i ~\pm~ 0.5\Ptwom ~+~ \DelPhi$), and a correction for 
scaling due to drift was applied; these were then mapped to a pair 
of $\eta$ values (for the specified geometry) and their difference 
converted to a \DelEta\/ value for that $\phi_i$.  Solutions that 
gave \DelEta\/ close to $360\degr/\Nsp\/$ across the entire pulse 
window were taken as acceptable solutions. The range of values of 
\Nsp, $k$ tried was 14,1 to 17,4 and for \Ap\/ and \Bp\/ it was 
0 to 2.2 degrees and 1.0 to 7.0 degrees, respectively.

The results from this analysis are summarized in Table \ref{tbl2},
which lists various parameters for the different solutions.
Acceptable solutions could be found for each choice of \Nsp.  
Of these, physically meaningful solutions (see section
\ref{sec:discuss} for a detailed explanation of this) are those
for $\Nsp\/ = 14,15$ and $16$, and only these are included in
Table \ref{tbl2}. From column 2 of this table, it is seen that 
as the number of sparks is increased, the solution requires a 
higher value of alias order $k$ (i.e. a larger value of drift 
rate). This is because the smaller \Ptwot\/ from a larger \Nsp\/ 
requires a faster drift so that the \Ptwot\/ values can be 
``stretched'' to match the range of observed \Ptwom\/ values. 
For every such combination of $\Nsp,k$ values, there is small 
range of \Ap,\Bp\/ values that result in equally acceptable 
solutions, in that they yield curves for \DelEta\/ versus 
longitude that are very similar, and have a mean value, 
$\DelEta_{avg}$, that is close to expected values 
$\Delta\eta_{exp}=360\degr/\Nsp$. This range is \Ap,\Bp\/ $\sim$ 
0\fdg4,2\fdg0 to 1\fdg2,5\fdg2 with $\Ap < \Bp$ always -- 
this is in contrast to the case of the dipole solutions obtained 
in section \ref{sec:geometry}, where $\A > \B$.  However, it is 
quite interesting to note that given the requirement 
$\Ap + \Bp = \A + \B$, it is possible to find a combination of 
\Ap\/ and \Bp\/ to match almost the entire range of dipole solutions 
in section \ref{sec:geometry}.  Column 3 of table \ref{tbl2} lists 
a typical set of \Ap,\Bp\/ values.  The middle panel of figure 
\ref{fig:P2_varn} shows the results for the \DelEta\/ values 
obtained for these \Ap, \Bp\/ combinations for case of the three 
solutions, with \DelPhi\/ set to $18\degr$.  The mean value of 
\DelEta\/, computed from all the points across the pulse window 
(except the 4 outliers), is shown in column 5 of table \ref{tbl2}. 
As can be seen, these $\DelEta_{avg}$ values are very close to the 
expected values of $360\degr/\Nsp$ (given in column 4 of the table). 
These results are very sensitive to small changes in values of \Ap 
and \Bp\/, with changes as small as $\sim 0\fdg2$ (in any one of 
the angles) leading to substantial deviations in the \DelEta\/ curves.  
The solutions are also quite sensitive to the value of \DelPhi, with 
small changes of the order of 10-15 degrees causing significant 
deviations from acceptable solutions.  This is illustrated with the 
three sample results in the bottom panel in fig \ref{fig:P2_varn}, 
which are computed for 3 different values of \DelPhi. The best results 
are obtained for $\DelPhi = 18\degr$, and the values in table 
\ref{tbl2} are for this choice.

The solutions for \Pthree\ and \Pfour\ are given in columns 6 and
7 of table \ref{tbl2}, both in units of seconds and in units of
the period, \Pone.  The results for the drift rate are in column
8. The percentage change in the amplitude of the drift rate for
the largest typical deviation mentioned in section \ref{sec:obs}
is given in column 9.  The largest fractional change of 7.4\% is
needed for the case of $\Nsp,k = 14,1$, and it reduces to 2.8\%
for the case of $\Nsp,k = 16,3$.  Thus, the changes in the mean
drift rate required to explain the observed drift rate variations
are only a few percent.

\section{Discussion and Summary}        \label{sec:discuss}

\paragraph{1. On the location and arrangement of sparks in PSR B0826$-$34 :}
Our analysis and interpretation of the data for \OurPsr\/ reveals
that we are looking at a pulsar that is very close to being an
aligned rotator, with a value of \A\/ less than 5\degr, and more
likely to be around 1\degr\/ to 2\degr.  Consequently, our line of
sight traces out a circular track centred almost around the dipole 
magnetic axis.  This not only gives the exceptionally wide profile 
that we observe for this pulsar, but also allows us to sample the 
emission pattern along a ring around the magnetic axis.  This radiation 
pattern reveals the presence of multiple drifting bands of subpulses 
(up to 7 are visible), which we interpret as produced by subpulse
producing plasma columns associated with multiple sparks circulating 
in the ring.  This is the first pulsar for which {\it direct} evidence 
for such a large number of sparks has been obtained -- all previous 
cases show evidence for at most 2-3 drift bands visible within the 
pulse window.  Hence our results and interpretation can be considered 
as another direct confirmation of the idea of rings of circulating 
sparks on the polar cap, which on the average form conal beams of 
emission \citep[][]{gks93,KG2002}.

The other interesting result is that the measured variation of
\Ptwo\/ -- the longitude separation between the multiple
drift-bands -- across the main pulse window is incommensurate with
the values of \A\/ and \B\/ for the dipole geometry that we infer
from the variation of main pulse width with radio frequency and
the known polarization angle variation.  This leads us to a new
and interesting conclusion : that the pattern of drifting
subpulses does not circulate around the dipole magnetic axis. 
The actual point of circulation (which we refer to as the 
``local pole''), is determined to be only $\sim 3\degr\/$ away 
from the dipole axis; however, this small difference is enough
to cause significant differences in the observed values of \Ptwo.

What determines the location of this centre of circulation and how
do we understand this in the context of the existing theoretical
models?  In our interpretation, this centre of circulation of the
drift pattern is a local pole produced by the more complicated
magnetic field structure that could exist close to the neutron
star surface.  Though we believe that the neutron star magnetic
field is practically dipolar at the heights at which the radio
emission is generated in the magnetosphere, there is now a growing
set of evidence that the magnetic field close to the neutron star
surface (up to a few stellar radii) may have higher order
multi-pole structure \citep[see ][ for review]{ug03}.  In
fact, it could be a ``sun-spot'' like magnetic field (e.g. GS00).
It is worth noting that the higher magnetic field strength and the
much smaller radii of curvature that such a field configuration
provides are actually essential for the working of the basic pair
production mechanism inside the vacuum gap that drives the RS75
kind of models \citep[][ and references therein]{gm02}. In such a
scenario, it is possible to have on the polar cap, a local pole of
the magnetic field configuration whose location is not aligned
with that of the pole of the dipolar field.  In models with vacuum
gaps close to the neutron star surface, it will be the structure of 
this non-dipolar magnetic field that will control the activities in 
the vacuum gap. In the model of GS00, the location of this local 
pole is taken as the location of the central spark (anchored to the 
local pole) that constitutes the core beam of emission. The rest of 
the polar cap is then populated by spark discharges that are arranged 
in concentric circles around the central spark, leading to a nested 
cone structure for the emission beam.  Within each circle, the sparks 
circulate due to the \ExB\ drift in a manner similar to the original 
RS75 model. Though the higher order field dominates close to the 
neutron star surface, with increasing height it is thought to make 
a smooth transition to the dipolar field that dominates at higher 
altitudes.  Thus we have a situation where the basic distribution of 
the seeds of emission (the sparks) on the polar cap is decided by 
the higher order surface magnetic field, while the actual height of 
the radio emission (and the evolution with frequency thereof) is 
dictated by the dipolar field at altitudes of typically about 50 
stellar radii.

Our results and inferences thus lend strong support to this picture of
strong non-dipolar magnetic fields close to the neutron star surface.

\paragraph{2. The two cone model for PSR B0826$-$34 :}
For most of the work presented here, we have concentrated on the emission
properties of the main pulse region of this pulsar.  However, as we noted in
section \ref{sec:obs}, there is a weak interpulse seen in our 318 MHz average
profile, which becomes gradually stronger as we go to higher frequencies,
and even dominates over the main pulse emission at frequencies higher than
$\sim$ 1 GHz (see fig. \ref{fig:freq_evln}).  The natural explanation for this
is emission from a second, outer conal ring with the geometry contriving such
that the line of sight samples a large fraction of the inner ring at the lower
frequencies ($\sim$ 300 MHz), while missing most of the outer ring.  Then, as
the emission beams reduce in angular size at the higher frequencies, the
line of sight first comes closer to the inner ring (in the main pulse region)
-- at around 600 MHz -- and then actually crosses beyond the inner ring at
frequencies above 1000 MHz, leading to relatively weak emission in the main
pulse window.  This evolution naturally explains the increased ``filling-up''
of the bridge region in the main pulse window as we go from 318 MHz to 645 MHz.
At the same time, the frequency evolution of the beams brings the outer conal
ring closer to the line of sight in the interpulse region.  This evolution is 
such that at around 600 MHz, significant part of the outer ring begins to be 
seen by the line of sight and at frequencies around 1.4 GHz, the line of sight 
is actually grazing along the outer ring in the interpulse region.

We have modeled this configuration of two conal rings using a simulation 
program that takes a given distribution of sparks on the polar cap and 
translates them along dipolar field lines to frequency dependent heights 
of emission in the magnetosphere and generates the observed pulse profiles 
that would be seen along a specified line of sight (see Appendix~B for 
details). The dashed lines in figure \ref{fig:freq_evln} show the simulated 
profiles obtained for a model having two conal rings of emission with 
frequency dependent radii determined by eqn.~(\ref{eqn:r_em}) and centred
around the local pole which is located at the point of the polar cap 
characterized by \Ap$=0\fdg436$, \Bp$=3\fdg064$ and $\Phi_{off}=17\fdg3$.
As can be seen, the match with the observed profiles is quite good, providing 
direct support for the model of 2 conal rings.  One important point worth 
stressing is that once the amplitude ratio of main pulse to interpulse is 
fixed at one frequency, then the spectral evolution of these profile 
components is almost entirely due to dipolar spreading of magnetic field 
lines in the emission region.

Our two conal ring model for \OurPsr\/ makes clear predictions about how
the average profile and single pulse characteristics should evolve with
frequency and suitable experiments can be devised for testing these.  For
example, one of the interesting things that it predicts is that the position
angle curve should be shifted with respect to the centre of the main pulse
window, because of the \DelPhi\/ rotational phase difference between the
fiducial planes containing the local pole axis and the dipole axis.  In
figure 6 of \cite {Biggs}, there is some hint that the centre of the PA
swing occurs at slightly later longitudes than the centroid of the main
pulse, but the quality of the data is not good enough to ascertain this.
Clearly, good quality polarimetric observations of \OurPsr\/ can help in
testing this aspect of our model.  Further, since our model predicts the
interpulse emission region to arise from a different conal ring, we expect
that the subpulse properties here could be different from those for the main
pulse region.  For example, depending on the number of sparks populating the
outer ring, the measured value of \Ptwo\/ could be very different from the
value we measure for the inner ring.  Perhaps the most interesting feature
would be the value of drift rate and inferred values of \Pthree\/ and \Pfour\/
in the interpulse region at 600 to 1400 MHz.

\paragraph{3. Pulse profile classification of PSR B0826$-$34 :}
It is clear from our work that there is a need for a revision of
the earlier classification of \OurPsr as a ``M-type'' pulsar 
\citep[e.g.][] {JMR93,LM88}, i.e. a pulsar where the line-of-sight 
and emission geometry are such that multiple cones of emission are 
visible along with the central core component.  In our model, all the
emission seen in the main pulse window originates from a single
conal ring of emission and that seen in the interpulse regions
comes from the second, outer conal ring of emission; the line of
sight thus does not sample the central core region at all.  Though
the average profiles of this pulsar sometimes tend to show partial
evidence for discrete emission components in the saddle region
between the leading and trailing emission peaks in the main pulse
region (as can be seen in figure \ref{fig:freq_evln}), we believe
these are artefacts of the quasi-longitude stationary drift
pattern of the multiple drift bands, especially when average
profiles are made over relatively short intervals of time, from
limited number of pulses. 

The classification of PSR B0826-34 as a ``M-type'' profile pulsar 
seemed strongly supported by the prominent circular polarization, 
reversing sense in the central part of the profile \citep[]
[their fig.~6]{Biggs}.  It therefore follows from our analysis of 
drifting subpulse patterns in this pulsar, that sense reversing 
circular polarization does not have to be associated with core 
emission only.  It can also appear in conal components, if the 
related subpulses (as in fig. \ref{fig:stack_grey}) are quasi-stationary 
in phase \citep[for more details see][]{gkms95}.

\paragraph{4. Inferences from the drift behaviour of PSR B0826$-$34 :}
We now discuss the results for the drifting behaviour of this pulsar
in some detail.  Not only do we see subpulse emission from several 
spark-associated plasma columns, we are also able to detect the presence 
of correlated drift pattern from all of these subpulses.  The observed 
apparent reversals of drift direction, which are difficult to explain 
under any physical model of the subpulse drift phenomenon, are naturally 
explained as a combination of aliasing of the true drift rate coupled 
with small ($\approx 3-8\%$) variations of the true drift rate.  Thus, in 
our interpretation, the intrinsic drift rate does not show any reversals 
in sign, but only small changes in amplitude, making the behaviour of 
this pulsar compatible with these models.  The inferred values of both 
the mean drift rate and the proposed small variations are worthy of a 
detailed discussion, which we now present.

As we have shown, a unique solution for the spark and emission
geometries for \OurPsr\/ is not possible with the current data, as
the alias order and \Nsp\/ can not be unambiguously determined.
Instead, for a given choice of the value of one of these, the
value for the other that yields acceptable solutions can be found.
The values for \D, \Pthree\/ and \Pfour\/ can then be found for these
cases and compared with expectations from models.  These results are
summarised in Table \ref{tbl2}, for three choices of the value of
\Nsp. The drift rates for most of the solutions are much lower
than the values expected for the pure vacuum gap models (e.g. RS75,
GMG03).  Correspondingly, the values of \Pfour\/ are
substantially longer than the prediction.  This result for
\OurPsr\/ is similar to the results obtained for PSR B0943+10
\citep {DR99} and PSR B0809+74 \citep {vL_etal_2003}. A natural
explanation for this behaviour is provided by GMG03 who
show that if the \ExB\/ drift is considered in a non-ideal vacuum
gap, i.e. one in which there is a finite amount of charge flow
\citep [e.g.][]{CR80}, then the predicted value of the drift rate
is reduced.  The proposed agency for the control of the ion flow
in the gap is thermal regulation, i.e. the ratio of the surface
temperature to the ion critical temperature controls the rate of
ion flow, and hence the electrical potential drop across the gap
and hence the value for the \ExB\/ drift rate.  This effect is
characterised by a ``shielding factor'' by GMG03 who find typical
values for it to be $\sim$ 0.2.  For our possible solutions, the
values of the shielding factor (SF) are shown in column 10 of
Table \ref{tbl2}.  For the slowest drift rate solution, this factor
is 0.36 and it increases to 0.94 (i.e. close to vacuum gap conditions)
for the case of $\Nsp,k = 16,3$.  From this it is clear that
solutions for $k>3$ are not viable as the implied drift rate would
be too fast -- more than the maximum that is produced under pure
vacuum gap conditions.

In the above scenario, the small variations around the mean drift
rate would then be due to small changes in the neutron star's
surface temperature in the polar cap region.  For example, to
produce a 8\% change in the drift rate, we find, using the
formulation in GMG03, that only a 0.14\% change in the surface
temperature is needed.  This means a change of about $3500$~K out
of a mean temperature of $\sim 2.5\times10^6$ K.  What could case
such small and cyclic (though quasi-periodic at best) variations
in the polar cap temperature is an interesting and open question.
One possibility is a kind of ``self-regulating'' mechanism of
heating and cooling of the polar cap by increased and reduced
spark activities. Whatever it maybe, from our earlier deductions
(at the end of section \ref{sec:aliasing})  about the regions of
slower and faster than normal drift rates, it would imply that the
process of cooling below the mean temperature (which would lead to
a faster than mean drift rate) is observed to be a relatively
slower and smoother process than that the reverse process of
heating (associated with slower than mean drift rates), which is
found to be more abrupt and faster. Further, it is interesting
that the observed drift rate fluctuations are found to be highly
correlated for all the visible drift bands (section \ref{sec:obs}), 
implying that the small temperature variations (if they are the 
cause of the drift fluctuations) occur simultaneously on a fairly 
large scale -- across a large part of the polar cap.  Here, 
detection and comparison of subpulse drift (and its variations) 
in the interpulse region (which we believe is due a second ring 
of sparks on the polar cap) with that in the main pulse region 
should produce valuable additional insight.

\paragraph{5. General implications for the average pulsar population:}
Finally, we turn to a discussion of the implications of slow
variations of drift rate for the average pulsar population.  If
the few percent variation of drift rate that we infer for
\OurPsr\/ is a generic phenomenon affecting a large number of
pulsars, then it could explain some of the mysteries of
non-detection of periodic drift variations in the fluctuation
spectra analysis of several pulsars.  Depending on the time scale
of the drift rate fluctuations, compared to the pulsar period and
the length of data taken for fluctuation spectra analysis, it
is possible that drift rate variations could significantly, or 
even completely, blur the periodic signal.  Thus it is possible
that several pulsars have drifting subpulses, but with some
fluctuations in the drift rate, leading to reduced chances of
detection of the phenomenon in a traditional fluctuation spectrum
analysis.  Even amongst the ``classical'' drifting pulsars (e.g.
PSR B0031$-$07, PSR B0943$+$10), variations in drift rate are
a common observed feature.

As a specific example, we would like to point out the case of PSR B0540+23,
which almost certainly shows stroboscopic phenomena due to aliasing, similar
to \OurPsr.  As reported by \citet{now91}, the subpulse drift is irregular,
without specified drift direction and with different speeds, including zero
speed. As a result, there is no specified value of $P_3$, and there is no
signature in the fluctuation spectrum.  Therefore, this pulsar is very similar
to PSR 0826-34 \citep[see Figs.~4, 5 and 6 in][]{now91}, except that the pulse
profile is much narrower, implying a larger inclination angle.

\paragraph{6. Summary:}  Using results from sensitive single pulse observations 
with the GMRT at 318 MHz, we have made significant progress in unraveling the
complex drifting subpulse pattern of \OurPsr.  In the process, we have shown
that this pulsar is very close to being an aligned rotator.  This explains the
unusually wide profile and multiple drift bands seen for this pulsar.  We have
modeled the observed profiles and shown evidence for the presence of two conal
rings of emission.  A detailed treatment of effects of aliasing and geometry on
the drifting subpulse pattern has been provided.  We have given an explanation 
for the observed variability of the drift rate, including the apparent reversals 
of drift direction, within the framework of the currently available theoretical 
models.  As a result, we have shown that small variations in the surface 
temperature of the neutron star polar cap can explain the observed drift pattern.  
Finally, we have also demonstrated that the subpulse pattern shows evidence for 
the presence of non-dipolar magnetic fields close to the neutron star surface.  
Clearly, there is a lot to be learnt from \OurPsr.

\begin{acknowledgements} 
We thank the staff of the GMRT for help with the 
observations.  The GMRT is run by the National Centre for Radio Astrophysics 
of the Tata Institute of Fundamental Research.  Y.G. would like to acknowledge 
the support from the Institute of Astronomy, University of Zielona Gora, for 
a short sabbatical visit during which this work was initiated. This paper is
supported in part by the Grant 2 P03D 008 19 of the Polish State Committee 
for Scientific Research. We thank E. Gil for technical help and R. Nityananda 
for a critical reading of the manuscript.
\end{acknowledgements} 

\appendix

\section{Pulsar Emission Geometry Formulae}

The geometry of pulsar radiation is described by \citet[][ see
their figure 10.4]{MT77}. At the rotational phase $\phi$, the
angle between the observer's line of sight and the dipole magnetic
axis is given by
\begin{equation}
\cos\Gamma ~~=~~ \cos\alpha\cos\zeta + \sin\alpha\sin\zeta\cos\phi ~~~,   \label{eqn:gamma}
\end{equation}
where $\zeta=\alpha+\beta$, $\alpha$ is the angle between the
rotation  and the magnetic axes and $\beta$ is the impact angle of
the closest approach of the line of sight to the magnetic axis.

The angle $\Gamma$ expressed above is also the angle between the
dipole axis and the tangent to magnetic field lines at points
where the emission observed at  the rotational phase $\phi$
originates. Using equations for the dipolar field lines, this
opening angle (also referred to as the beam radius) can be
described as
\begin{equation}
\rho ~~=~~ 1\fdg24 ~s~ [r_{\rm em}(\nu)/R]^{1/2} ~\Pone^{-1/2}~~~,  \label{eqn:rho}
\end{equation}
where $\Pone\/$ is the pulsar period, $r_{\rm em}(\nu)$ is the
emission altitude (at which radiation at a given frequency $\nu$
is generated) and $R=10^6~$ cm is the neutron star radius.  The
mapping parameter $0\leq s=d/r_p\leq 1$ is determined by the locus
of dipolar field lines on the polar cap ($s=0$ at the pole and
$s=1$ at the polar cap edge), where $d$ is the distance from the
magnetic axis to the field line on the polar cap corresponding to
a certain detail of the pulse profile 
and $r_p = 1.4 \times 10^4 \Pone^{-1/2}$ cm
is the canonical polar cap radius.

According to the generally accepted concept of radius-to-frequency 
mapping, higher frequencies are emitted at lower altitudes than 
lower frequencies.  The emission altitude can be described by 
\citep [][and references therein]{KG2}
\begin{equation}
r_{\rm em} ~~=~~ (40\pm 8) R ~\nu_{\rm GHz}^{-0.26 \pm 0.09}
\dot{P}^{0.07\pm 0.03}_{-15} \Pone^{0.37 \pm 0.05}~~~, \label{eqn:r_em}
\end{equation}
where $R=10^6$ cm, $\nu_{\rm GHz}$ is the observing frequency (in GHz), 
$\Pone\/$ is the pulsar period (in seconds) and $\dot{P}_{-15}$ is the 
period derivative in units of $10^{-15}$~s/s.

The polarization position angle (P.A.) curve is described by
\begin{equation}
tan\psi ~~=~~ \frac{\sin\alpha \sin\phi}{\sin\zeta \cos\alpha - \cos\zeta \cos\phi} ~~~;  \label{eqn:PA}
\end{equation}

and the value of the steepest gradient of the P.A. curve is given by

\begin{equation}
\frac {d\psi} {d\phi}|_{max} ~~=~~ \frac {\sin\alpha} {\sin\beta} ~~~. \label{eqn:gradient}
\end{equation}

At a given frequency, $\nu_{\rm GHz}$, for a chosen combination of 
$\alpha,\beta$ and $s$, the value of the beam radius, $\rho$, can be 
found using equations (\ref{eqn:r_em}) and (\ref{eqn:rho}).  Then,
using $\Gamma(\phi)=\rho(\nu)$, the profile width corresponding to 
the emission peaks of the conal components, $2\phi$, can be found from
eqn. (\ref{eqn:gamma}) and compared with observed values.
For a chosen combination of $\alpha,\beta$, the steepest gradient can 
be computed using eqn (\ref{eqn:gradient}) and compared with observations; 
further, using the estimate of $\phi$, the total swing of the polarization 
angle across the pulse, $\Delta\psi$, can be computed from eqn (\ref{eqn:PA})
and compared with observed data.  Combinations of $\alpha,\beta$ and $s$ that
meet all these criteria for all the frequencies listed in table \ref{tbl1}
were found by doing a grid search in the values of $\alpha,\beta$ and $s$.

\section{Simulations of aliasing effects}

To illustrate the effects of aliasing in the subpulse drift patterns 
of \OurPsr, we simulate the radiation of this pulsar, assuming that 
its single pulse structure reflects the circumferential motion of 14 
sparks at a distance $d=0.44r_p=33$~ meters from the pole. For the 
emission geometry we adopted $\alpha=2\fdg5$ and $\beta=1\fdg0$, 
and an emission altitude for the frequency of 318 MHz was estimated 
from eqn.~(\ref{eqn:r_em}). We use the simulation technique described 
in GS00 (and references therein).  Each spark is modeled as a circular 
entity with a characteristic dimension equal to the height $h$ of the 
quasi-steady acceleration gap formed above the polar cap. The adjacent 
sparks are separated from each other also by about $h$ (spark centers 
are separated by $2h$). Since the elementary pulsar radiation is 
relativistically beamed along dipolar magnetic field lines, the 
spark-associated intensity pattern can be transformed from the polar 
cap to the emission region and then to the observer according to a given 
geometry $(\alpha,\beta,r_{em})$. The spark-associated subbeams, related 
to subpulses observed in single pulses, perform a circumferential motion 
due to the \ExB\ drift of the spark plasma.  The simulation program allows 
for circulation around either the magnetic dipole or local pole of 
surface magnetic field (assuming that this surface field is also axially 
symmetric on the polar cap), with a desired angular drift rate, 
$D=360\degr/P_4$, where $P_4$ is the period of spark circulation.

For given values of $P,\ \dot{P},\ \alpha$ and $\beta$ the number of 
subpulses and their phases in a single pulse are determined by the 
value of $D_p$. We performed an experiment with time varying drift rates, 
starting with $D_p=0$ and incrementing it by $0\fdg03$ every rotational 
period $P_1$ (this increment was chosen to make sure that cycles of gradual
variations of the drift rate have duration of about $100P_1$, as
observed in real data). The idea was to find a sequence of about 100 pulses
with varying $D_p$ in a range appropriate to produce curved drift-bands 
similar to those visible in Fig.~\ref{fig:stack_grey}. A sample result
of our simulations is presented in Fig.~\ref{fig:simulations}, with all 
the relevant information provided in the top panel and on both sides 
of the pulse window. The last column indicates the actual value of 
$D_p=0\fdg03~n$ per period, where $n$ is the rotational period number 
indicated on the vertical axis. Next to $D_p$ we list values of 
$P_4/P_1=360\degr/D_p$. On the other side of the pulse window we give 
values of $P_3/P_1=P_4/(14P_1)$ and the column just before shows the 
fluctuation frequency, $f_3P_1=P_1/P_3$.

What can we learn from simulated patterns presented in Fig.~\ref{fig:simulations}?
First of all, pulses around $n=1$ show that were there no drift $(D_p=0)$, 
the observer would clearly see 7 out of 14 sparks in the form of longitude
stationary subpulses. As the drift rate increases with the increasing 
number of periods, the subpulse drift with time varying rate becomes more 
and more apparent. However, up to about $n=100$, the subpulse drift is 
relatively slow, non-aliased and proceeds from the leading to the trailing 
edge of the profile. This is the real drift direction and the observed 
drift-bands are formed by the same sparks. This is, however, not true in 
the region well above $n=100$, where all kinds of stroboscopic aliasing
effects become visible. We have marked regions where the apparent drift-bands 
are formed by subpulses appearing at approximately the same phase every 
$m$-th period, where $m=5,4,3,2$ and so on.  It is worth noting that the 
number of apparent drift-bands is about 7m. The drift-bands change the 
apparent drift direction due to the aliasing effect, every time $f_3$ crosses 
a multiple of the Nyquist frequency.

Obviously, the region below $n=800$ does not correspond to drifting 
subpulses in PSR 0826-34, because it shows alternating (even-odd), 
longitude stationary intensity modulations, which are not observed in 
this pulsar. It seems, however, that its drifting subpulse patterns are 
well modeled by the region between $n=800$ and $n=900$, which could
represent one cycle of multiple curved drift-bands visible in 
Fig.~\ref{fig:stack_grey}. A clear pattern of seven drift-bands is 
visible, moving in an aliased direction (from the trailing to the leading 
edge) in the first half of the cycle, and in the true direction (from the 
leading to the trailing edge) in the second half of the cycle. The drift 
direction change occurs at $f_3=1/P_1$ (or $P_3=P_1$), which is twice the 
conventional Nyquist frequency. At this stage the pattern advances exactly 
by one subbeam per one pulsar period $P_1$ and the apparent drift-bands 
are formed by successive adjacent subbeams. The corresponding values of 
$P_4/P_1=14$ and $D_p=25\fdg7$ per period.  However, 50 pulses earlier, 
at the beginning of a cycle, $P_4/P_1=14.4$ and $D_p=25\degr$ per period, 
while about 50 pulses later, at the end of a cycle, $P_4/P_1=13.3$ and 
$D_p=27\fdg1$ per period.  This means that the pattern speeds up along 
each cycle, increasing $D_p$ by about 8\%. 

Although we present simulations only for the case of 14 sparks, the 
qualitative results are very similar for other values of $N_{sp}$. Results 
of simulations for 15 and 16 sparks, as well as those for 14 sparks presented 
in Fig.~\ref{fig:simulations}, agree exactly with the drift rate solutions 
presented in Table~\ref{tbl2}.

\clearpage

\begin{figure*}
\centering
\includegraphics[width=17cm]{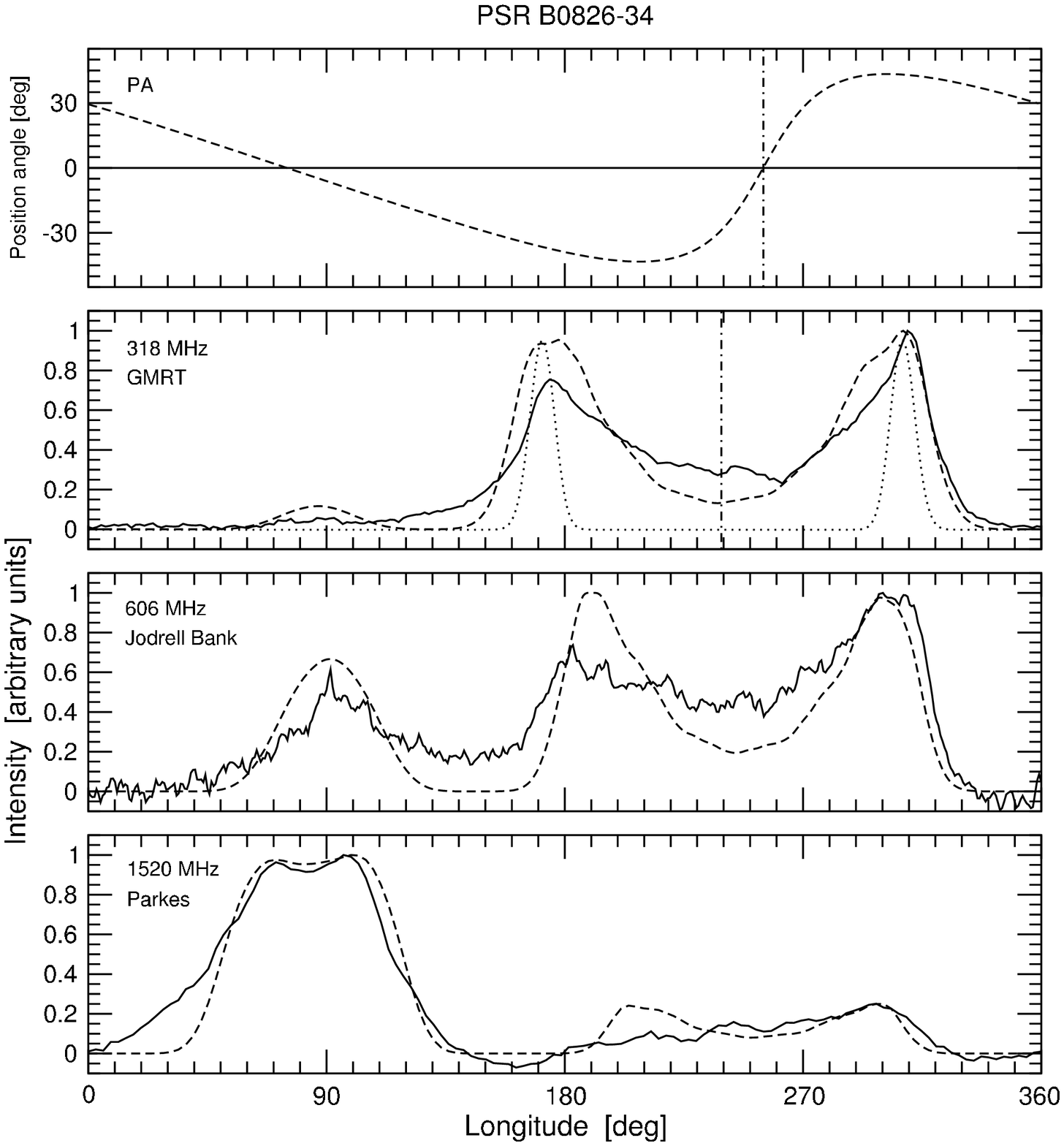}
\caption{Frequency evolution of the average profile of PSR B0826$-$34.  
The solid lines (bottom 3 panels) show the observed profiles at 3 different 
radio frequencies, including the result (at 318 MHz) from this paper.  
The dashed lines (bottom 3 panels) show the best fit simulated profiles 
corresponding to sparks circulating around the local surface magnetic pole, 
while the dotted line corresponds to spark circulation around the dipole 
axis (see text for explanations).  The top panel shows the position angle 
curve corresponding to dipolar geometry with $\alpha=2\fdg5$ and 
$\beta=1\fdg0$.  Note that the inflexion point of the position angle 
curve lags the midpoint of the main pulse by about $15\degr$.
\label{fig:freq_evln}}
\end{figure*}

\begin{figure*}
\centering
\includegraphics[width=17cm]{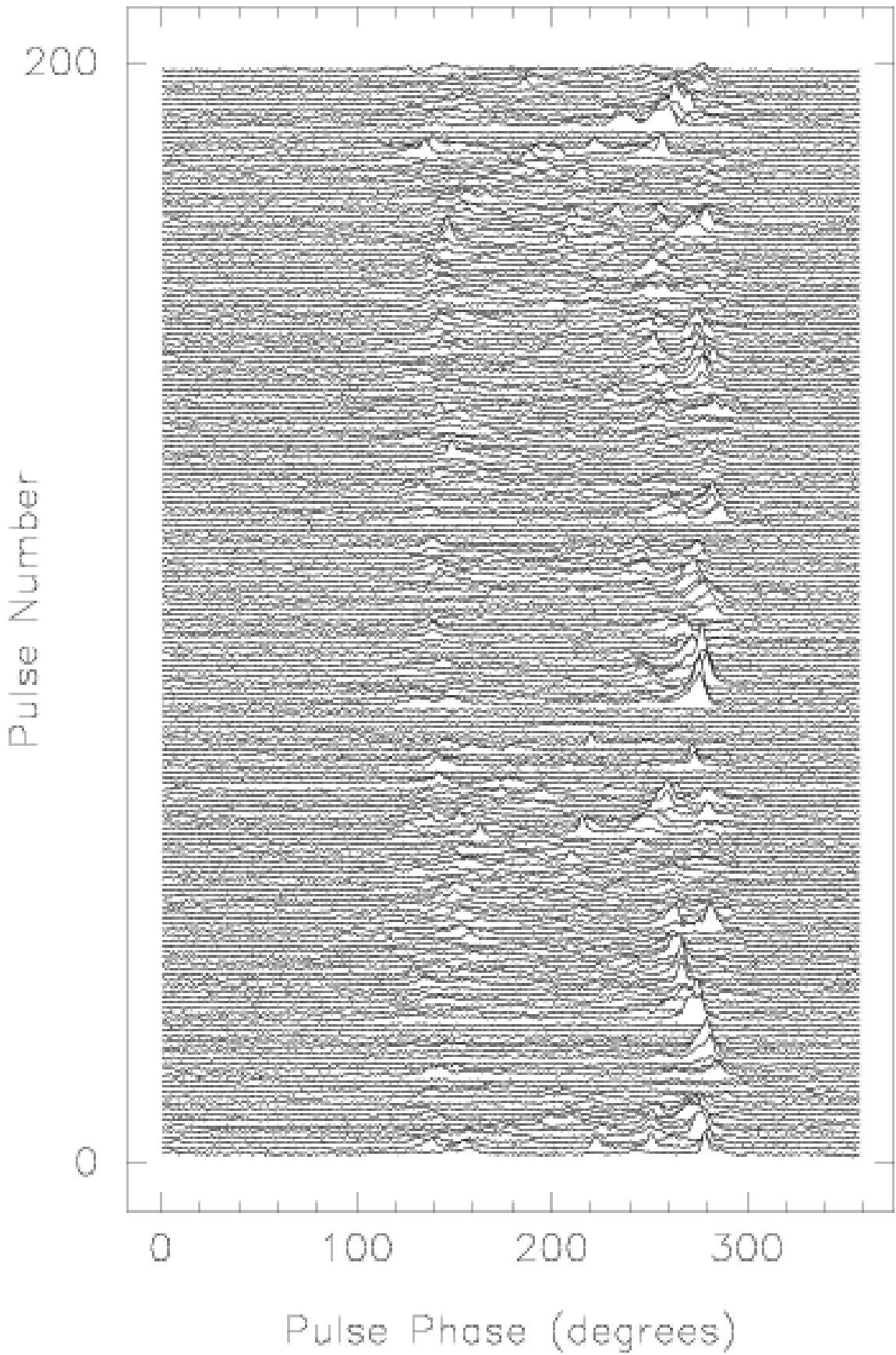}
\caption{Single pulse sequence of the first 200 pulses from the GMRT
observations of PSR B0826$-$34 at 318 MHz.
\label{fig:stack_line}}
\end{figure*}

\begin{figure*}
\centering
\includegraphics[width=17cm]{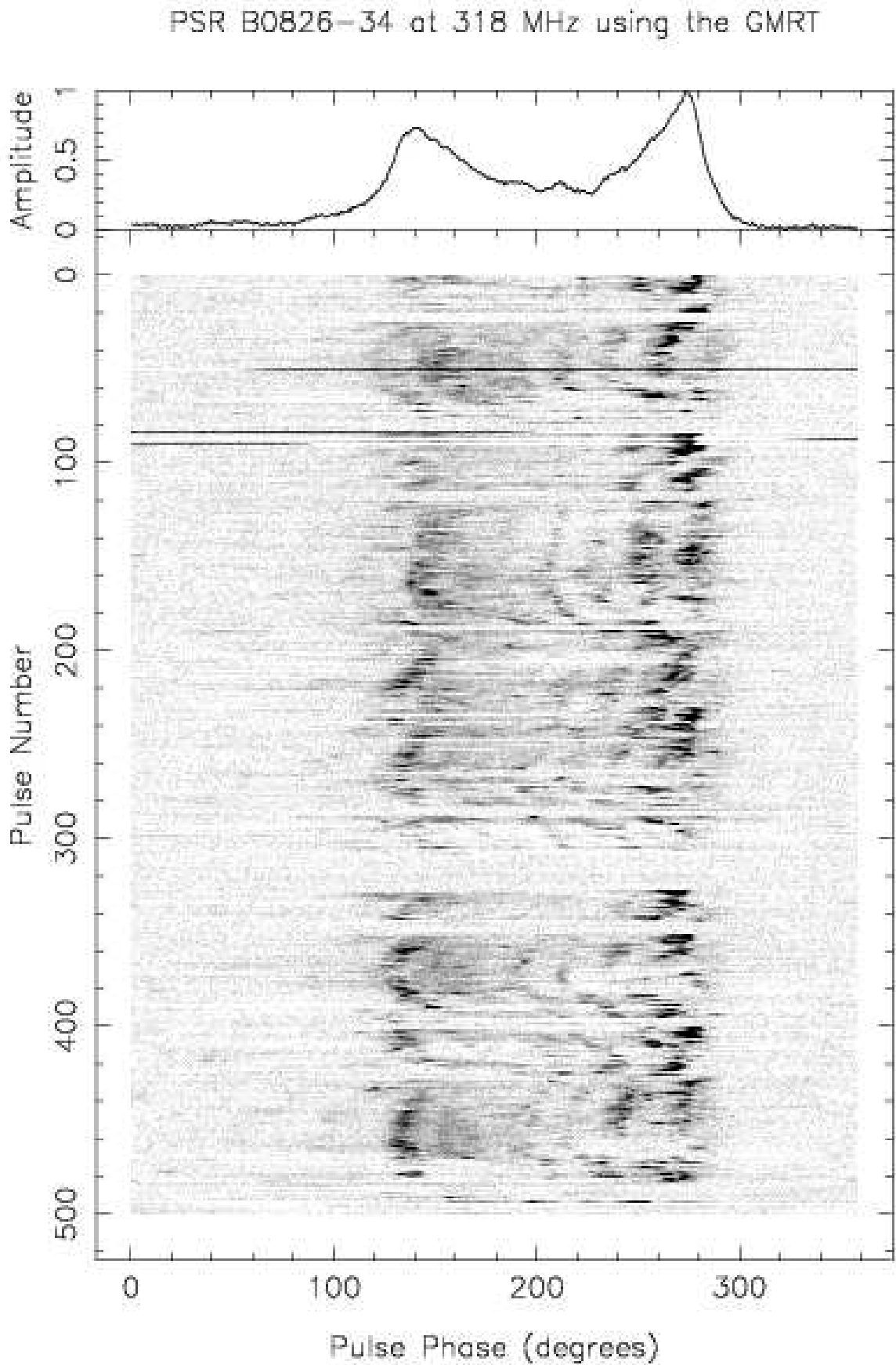}
\caption{Grey scale plot of single pulse data for 500 pulses of PSR B0826$-$34 
from the GMRT observations at 318 MHz, with the average profile shown on top.  
The dark horizontal lines between pulse numbers 50,60 and between 90,100 are 
due to radio frequency interference. 
\label{fig:stack_grey}}
\end{figure*}

\begin{figure*}
\centering
\includegraphics[width=8cm]{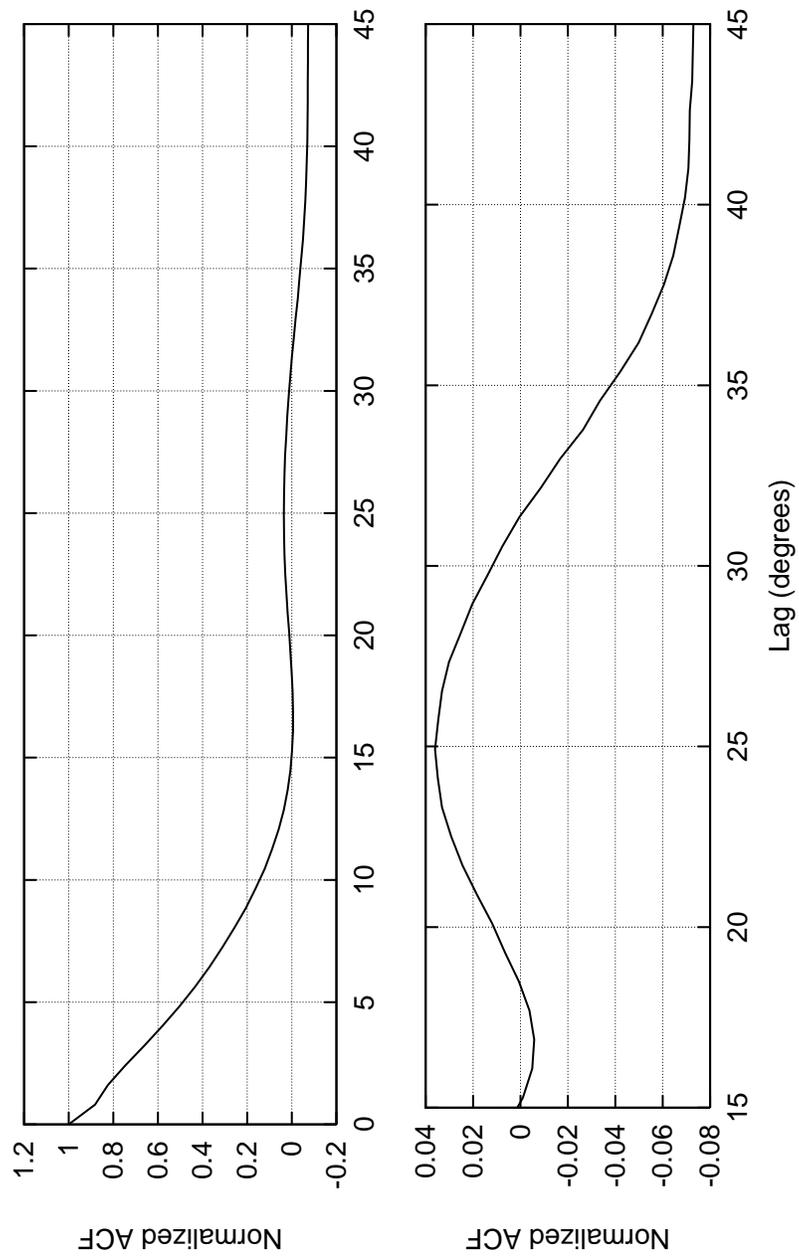}
\caption{Autocorrelation results for the entire on-pulse window data in figure
\ref {fig:stack_grey}.  The lower panel shows a detailed view around the secondary
maxima (near the lag of 25 degrees) of the autocorrelation function.
\label{fig:full_acf}}
\end{figure*}

\begin{figure*}
\centering
\includegraphics[width=17cm]{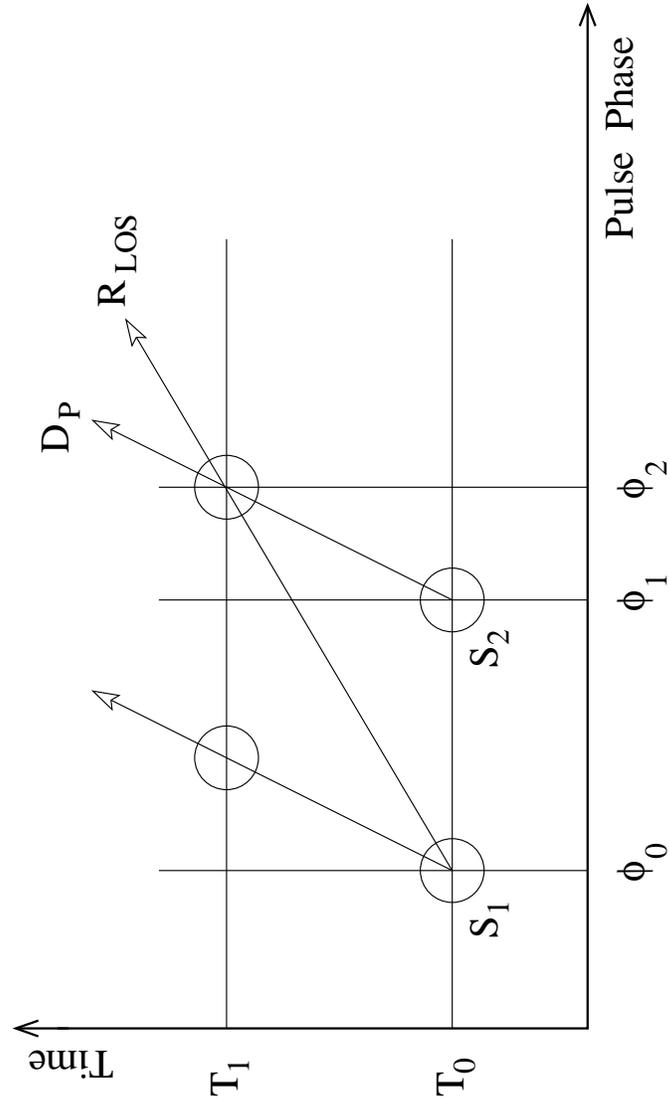}
\caption{Illustration of the concept of drift induced modification of \Ptwo.  
$S_1$ and $S_2$ are two sparks located at longitudes $\phi_0$ and $\phi_1$ at 
time $T_0$ (i.e. $\Ptwot = \phi_1 - \phi_0$).  At $T_0$, the line of sight 
(moving along $R_{LOS}$) is at $\phi_0$, and sees $S_1$.  At a later time 
$T_1$, the line of sight is at $\phi_2$, where it sees spark $S_2$.  Hence 
$\Ptwom = \phi_2 - \phi_0$. During the interval $T_1 - T_0$, the sparks
(moving along $D_P$) drift by an amount $\phi_2 - \phi_1$.
\label{fig:P2_stretch}}
\end{figure*}

\begin{figure*}
\centering
\includegraphics[width=13cm]{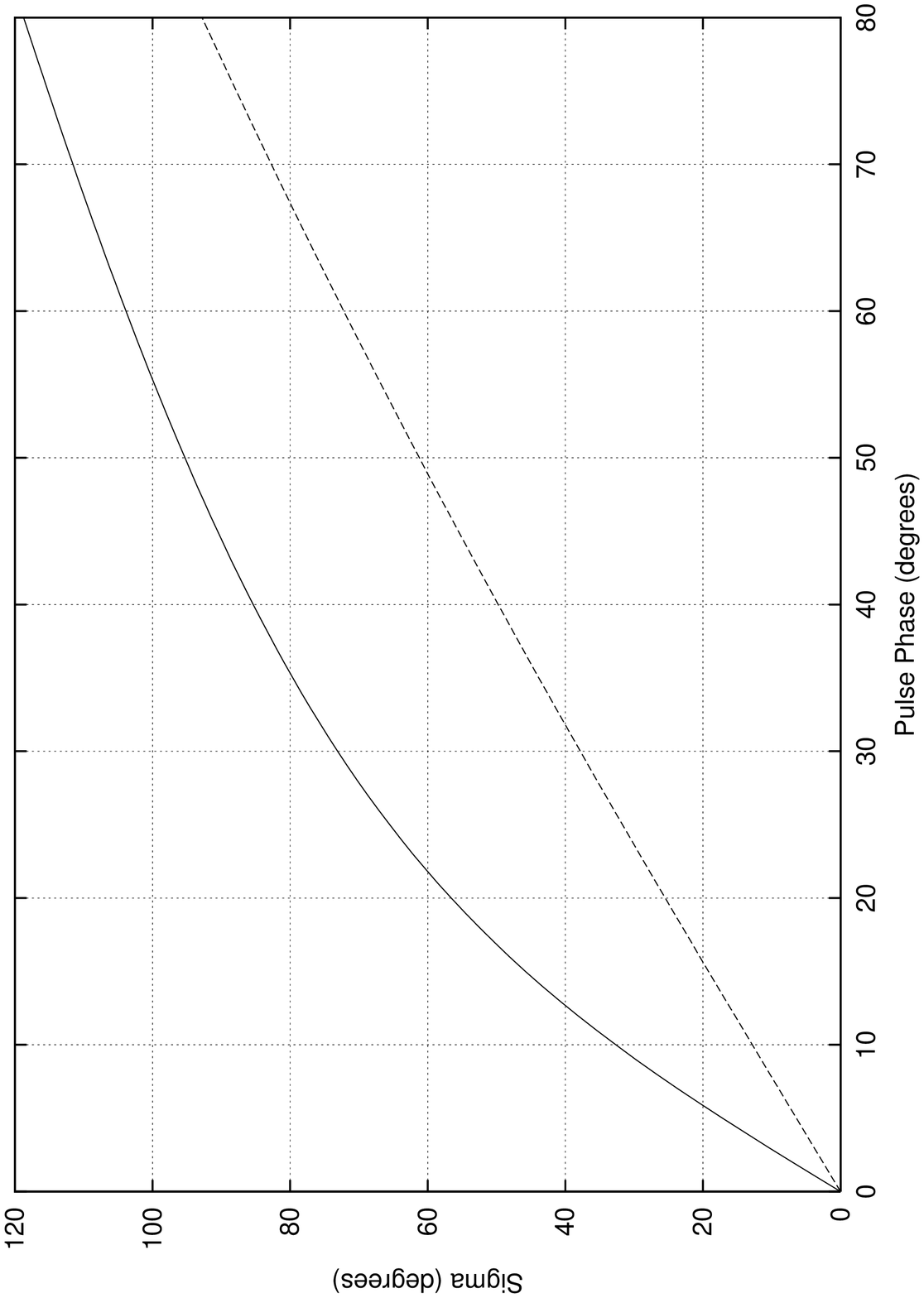}
\caption{Variation of $\sigma$ with $\phi$ (see eqn. \ref{eqn:sigma-phi} of 
text) for two different choices of geometry: $\A,\B = 2\fdg5,\ 1\fdg0$ 
(solid curve) and $\A,\B = 0\fdg8,2\fdg8$ (dashed curve). The phase
$\phi=0\degr$ corresponds to the fiducial phase.
\label{fig:sigma_plot}}
\end{figure*}

\begin{figure*}
\centering
\includegraphics[width=13cm]{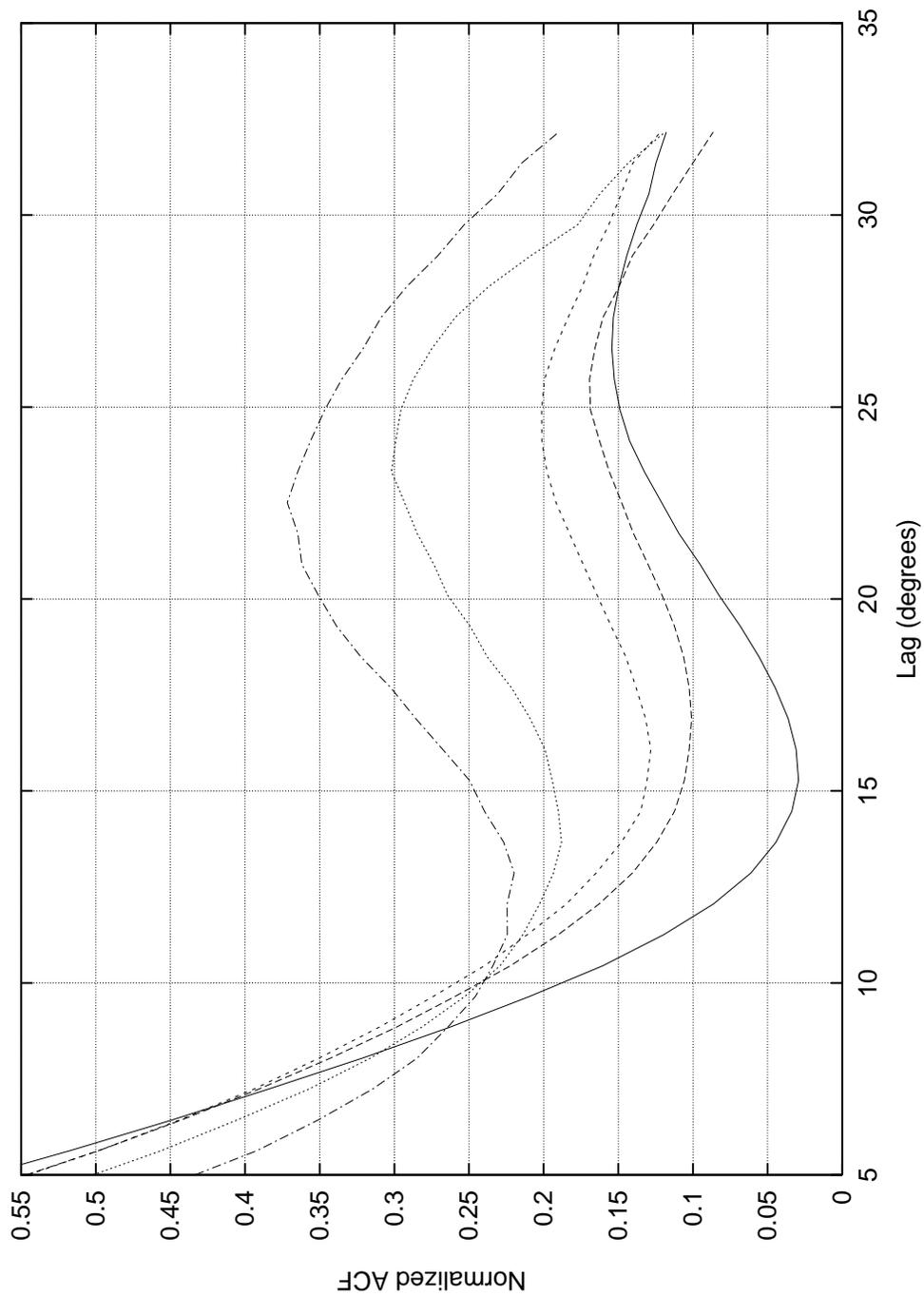}
\caption{Sample results of the autocorrelation function for narrow pulse longitude
windows of the data in figure \ref {fig:stack_grey}.  Curves for 5 different window
locations, centred at 269\degr~ (solid), 253\degr~ (long dashes), 245\degr~ (short 
dashes), 229\degr~ (dots) and 213\degr~ (dash-dot) of pulse longitude are shown to 
illustrate the variation of \Ptwom\/ with longitude.
\label{fig:window_corr}}
\end{figure*}

\begin{figure*}
\centering
\includegraphics[width=15cm]{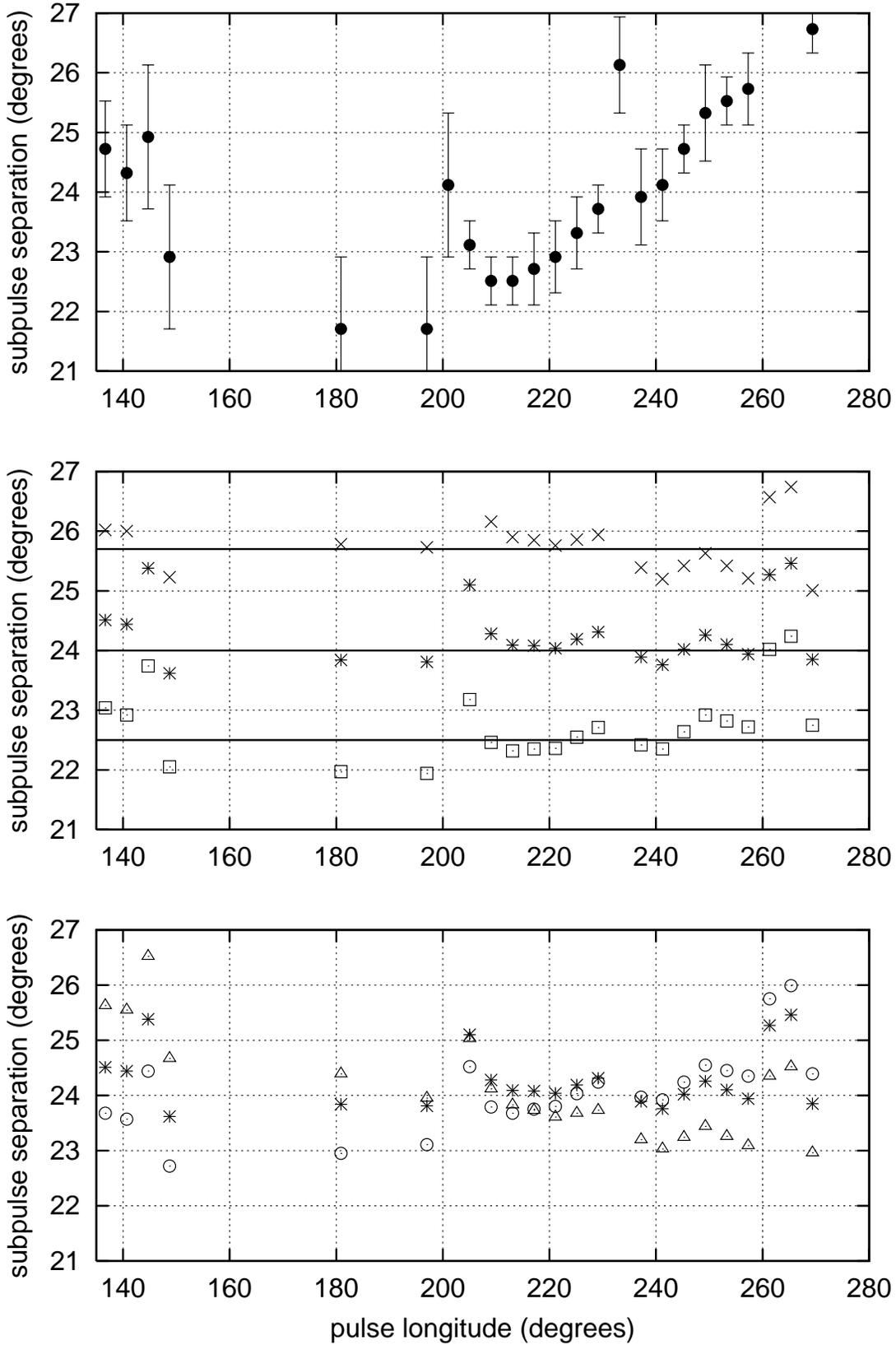}
\caption[width=10cm]{Top panel: Measured variation of \Ptwom\/ with longitude.  Middle panel: 
Inferred \DelEta\/ values as a function of pulse longitude, for each of the solutions 
list in table \ref{tbl2}; the 3 horizontal solid lines are the expected values 
($360\degr/\Nsp$) for each case.  Bottom panel: \DelEta\/ values for 3 cases of 
$\Nsp,k = 15,2$ solution, with $\DelPhi = 18\degr$ (asterisks), $\DelPhi = 28\degr$ 
(triangles) and $\DelPhi = 8\degr$ (circles).
\label{fig:P2_varn}}
\end{figure*}

\begin{figure*}
\centering
\includegraphics[width=15cm]{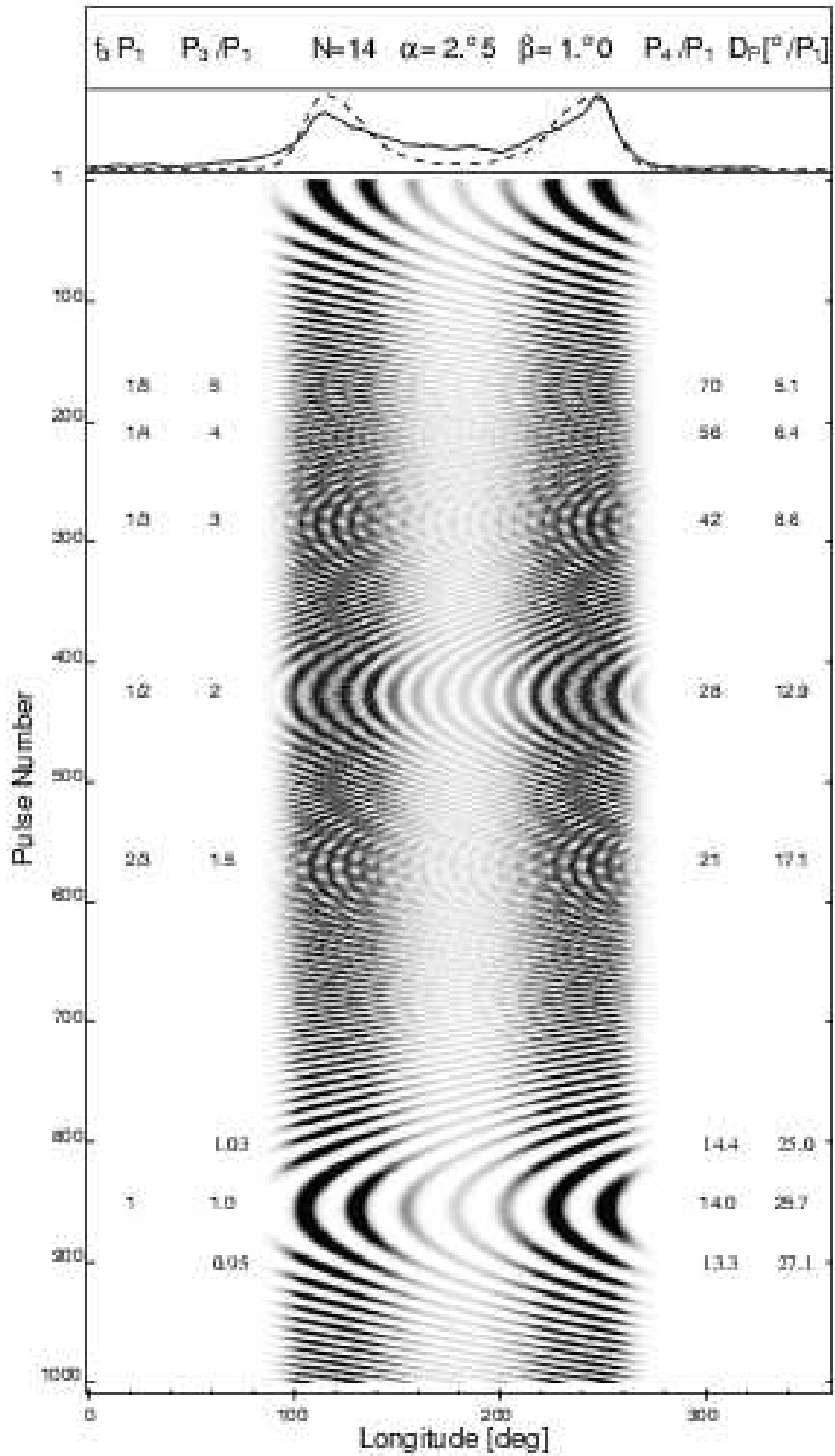}
\caption{Simulations of the subpulse drift phenomenon with time varying drift 
rate $D_p(n)=0\fdg03 n$ per period, where $n$ is the number of pulsar periods 
shown on the vertical axis (see text for explanation of other symbols). The 
quasi-periodic patterns of drifting subpulses observed in PSR B0826-34 
(Fig.~3) are well modeled by the sector between $n=800$ and $n=900$, with 
$D_p$ increasing from $25\degr$ to $27\fdg1$ per period.
\label{fig:simulations}}
\end{figure*}

\clearpage 

\begin{table}
\begin{center}
\caption[]{Frequency dependence of component separation.
\label{tbl1}}
\begin{tabular}{rcl} 
\hline
\noalign{\smallskip}
Frequency & Separation    &  References \\
(MHz)     & \rm (degrees) &             \\
\hline
\noalign{\smallskip}
\rm 610   & \rm 112 $\pm$ 4 & \rm \citet {Biggs}                   \\
\rm 408   & \rm 128 $\pm$ 4 & \rm \citet {Biggs}; \citet {Durdin}  \\
\rm 318   & \rm 134 $\pm$ 4 & \rm This paper                       \\
\rm 116   & \rm 151 $\pm$ 4 & \rm \citet {Turner}                  \\
\rm  95   & \rm 155 $\pm$ 4 & \rm \citet {Turner}                  \\
\rm  82   & \rm 157 $\pm$ 4 & \rm \citet {Turner}                  \\
\noalign{\smallskip}
\hline
\end{tabular}
\end{center}
\end{table}

\begin{table}
\begin{center}
\caption[]{Spark geometry and drift rate solutions for \OurPsr
\label{tbl2}}
\begin{tabular}{ccccccccccc}
\hline
\noalign{\smallskip}
\Nsp & $k$ & \Ap,\Bp & $\DelEta_{exp}$ & $\DelEta_{avg}$ & \Pthree & \Pfour & \Dp   &  $\Delta\Dp/\Dp$ &  SF  \\
     &     &(\degr) & (\degr)    & (\degr)    & (sec) (\Pone) & (sec) (\Pone) & (\degr/pulse) & (\%) & \\
\hline
\noalign{\smallskip}
 14 &  1 & \rm 0.8,2.8 & 25.7 & 25.64 & \rm 1.850 (1.00\Pone) & \rm 25.90 (14.0\Pone) & \rm 25.7 & \rm $<$ 7.4 & 0.36 \\
 15 &  2 & \rm 0.6,2.2 & 24.0 & 24.06 & \rm 0.925 (0.50\Pone) & \rm 21.95 (7.50\Pone) & \rm 48.0 & \rm $<$ 4.0 & 0.67 \\
 16 &  3 & \rm 0.8,3.2 & 22.5 & 22.52 & \rm 0.617 (0.33\Pone) & \rm ~9.86 (5.33\Pone) & \rm 67.5 & \rm $<$ 2.8 & 0.94 \\
\noalign{\smallskip}
\hline
\end{tabular}
\end{center}
\end{table}

\end{document}